\documentclass[twocolumn]{aastex63}
\usepackage{apjfonts}
\usepackage{amsmath}
\pdfpageattr{/Group <</S /Transparency /I true /CS /DeviceRGB>>} 
\begin{document}

\title{OGLE-ing the Magellanic System: Optical Reddening Maps of the Large and Small Magellanic Cloud from Red Clump Stars}

\author[0000-0001-9439-604X]{D.~M.~Skowron}
\affiliation{Astronomical Observatory, University of Warsaw, Aleje Ujazdowskie 4, \mbox{00-478} Warszawa, Poland}
\email{dszczyg@astrouw.edu.pl}
\author[0000-0002-2335-1730]{J.~Skowron}
\affiliation{Astronomical Observatory, University of Warsaw, Aleje Ujazdowskie 4, \mbox{00-478} Warszawa, Poland}
\author[0000-0001-5207-5619]{A.~Udalski}
\affiliation{Astronomical Observatory, University of Warsaw, Aleje Ujazdowskie 4, \mbox{00-478} Warszawa, Poland}
\author[0000-0002-0548-8995]{M.~K.~Szyma\'{n}ski}
\affiliation{Astronomical Observatory, University of Warsaw, Aleje Ujazdowskie 4, \mbox{00-478} Warszawa, Poland}
\author[0000-0002-7777-0842]{I.~Soszy\'{n}ski}
\affiliation{Astronomical Observatory, University of Warsaw, Aleje Ujazdowskie 4, \mbox{00-478} Warszawa, Poland}
\author[0000-0002-9658-6151]{\L.~Wyrzykowski}
\affiliation{Astronomical Observatory, University of Warsaw, Aleje Ujazdowskie 4, \mbox{00-478} Warszawa, Poland}
\author[0000-0001-6364-408X]{K.~Ulaczyk}
\affiliation{Astronomical Observatory, University of Warsaw, Aleje Ujazdowskie 4, \mbox{00-478} Warszawa, Poland}
\affiliation{Department of Physics, University of Warwick, Gibbet Hill Road, Coventry, CV4~7AL,~UK}
\author[0000-0002-9245-6368]{R.~Poleski}
\affiliation{Astronomical Observatory, University of Warsaw, Aleje Ujazdowskie 4, \mbox{00-478} Warszawa, Poland}
\author[0000-0002-9245-6368]{S.~Koz{\l}owski}
\affiliation{Astronomical Observatory, University of Warsaw, Aleje Ujazdowskie 4, \mbox{00-478} Warszawa, Poland}
\author[0000-0002-2339-5899]{P.~Pietrukowicz}
\affiliation{Astronomical Observatory, University of Warsaw, Aleje Ujazdowskie 4, \mbox{00-478} Warszawa, Poland}
\author[0000-0001-7016-1692]{P.~Mr\'{o}z}
\affiliation{Astronomical Observatory, University of Warsaw, Aleje Ujazdowskie 4, \mbox{00-478} Warszawa, Poland}
\affiliation{Division of Physics, Mathematics, and Astronomy, California Institute of Technology, Pasadena, CA 91125, USA}
\author[0000-0002-9326-9329]{K.~Rybicki}
\affiliation{Astronomical Observatory, University of Warsaw, Aleje Ujazdowskie 4, \mbox{00-478} Warszawa, Poland}
\author[0000-0002-6212-7221]{P.~Iwanek}
\affiliation{Astronomical Observatory, University of Warsaw, Aleje Ujazdowskie 4, \mbox{00-478} Warszawa, Poland}
\author[0000-0002-3051-274X]{M.~Wrona}
\affiliation{Astronomical Observatory, University of Warsaw, Aleje Ujazdowskie 4, \mbox{00-478} Warszawa, Poland}
\author[0000-0002-1650-1518]{M.~Gromadzki}
\affiliation{Astronomical Observatory, University of Warsaw, Aleje Ujazdowskie 4, \mbox{00-478} Warszawa, Poland}

\begin{abstract}

\noindent 
We present the most extensive and detailed reddening maps of the Magellanic
Clouds (MCs) derived from the color properties of Red Clump (RC) stars.
The analysis is based on the deep photometric maps from the fourth phase of
the Optical Gravitational Lensing Experiment (OGLE-IV), covering approximately
670~deg$^2$ of the sky in the Magellanic System region.
The resulting maps provide reddening information for 180~deg$^2$ in the Large
Magellanic Cloud (LMC) and 75~deg$^2$ in the Small Magellanic Cloud (SMC),
with a resolution of $1.7' \times 1.7'$ in the central parts of the MCs,
decreasing to approximately $27' \times 27'$ in the outskirts.
The mean reddening is $E(V-I\,) = 0.100 \pm 0.043$~mag in the LMC and
$E(V-I\,) = 0.047 \pm 0.025$~mag in the SMC.

We refine methods of calculating the RC color to obtain the highest possible
accuracy of reddening maps based on RC stars. Using spectroscopy of red
giants, we find the metallicity gradient in both MCs, which causes a slight
decrease of the intrinsic RC color with distance from the galaxy center
of $\sim0.002$ mag/deg in the LMC and between $0.003$ and $0.009$ mag/deg
in the SMC. The central values of the intrinsic RC color are $0.886$ and
$0.877$~mag in the LMC and SMC, respectively.

The reddening map of the MCs is available on-line both in the downloadable
form and as an interactive interface.

\end{abstract}

\keywords{galaxies: Magellanic Clouds -- interstellar medium: dust, extinction 
-- stars: general -- stars: statistics -- surveys: OGLE}

\section{Introduction}
\label{sec:introduction}

The Large and the Small Magellanic Clouds (LMC and SMC, respectively) are
a pair of irregular galaxies that due to their proximity to the Milky Way and
their mutual interactions are the most often studied nearby galaxies. They
serve as a local laboratory for investigating galaxy structure, evolution
and interactions, as well as stellar populations and interstellar medium
distribution. The Magellanic Clouds (MCs) are also used to calibrate the
cosmological distance scale.

The majority of stellar population studies require accounting for interstellar
extinction in order to acquire precise distances, thus it is very important to
have accurate reddening maps of the Magellanic System. Previous efforts to
construct reddening maps employed various tracers such as Red Clump stars
(\citealt{Udalski1999a,Udalski1999b}, \citealt{Subramaniam2005},
\citealt{Subramanian2009,Subramanian2013}, \citealt{Haschke2011}, 
\citealt{Tatton2013}, \citealt{Choi2018}, \citealt{Gorski2020}),
classical Cepheids (\citealt{Inno2016}, \citealt{Joshi2019}) or RR Lyrae type
variables (\citealt{Pejcha2009}, \citealt{Haschke2011}, \citealt{Deb2017}).
Interstellar reddening has also been inferred from equivalent widths
of spectral lines \citep{Munari1997},  stellar atmosphere fitting
(\citealt{Zaritsky2002,Zaritsky2004}), and through the analysis of
spectral energy distributions of background galaxies (\citealt{Bell2019}).

\vspace{0.1cm}
\begin{figure*}[ht]
\centerline{\includegraphics[width=0.95\textwidth]{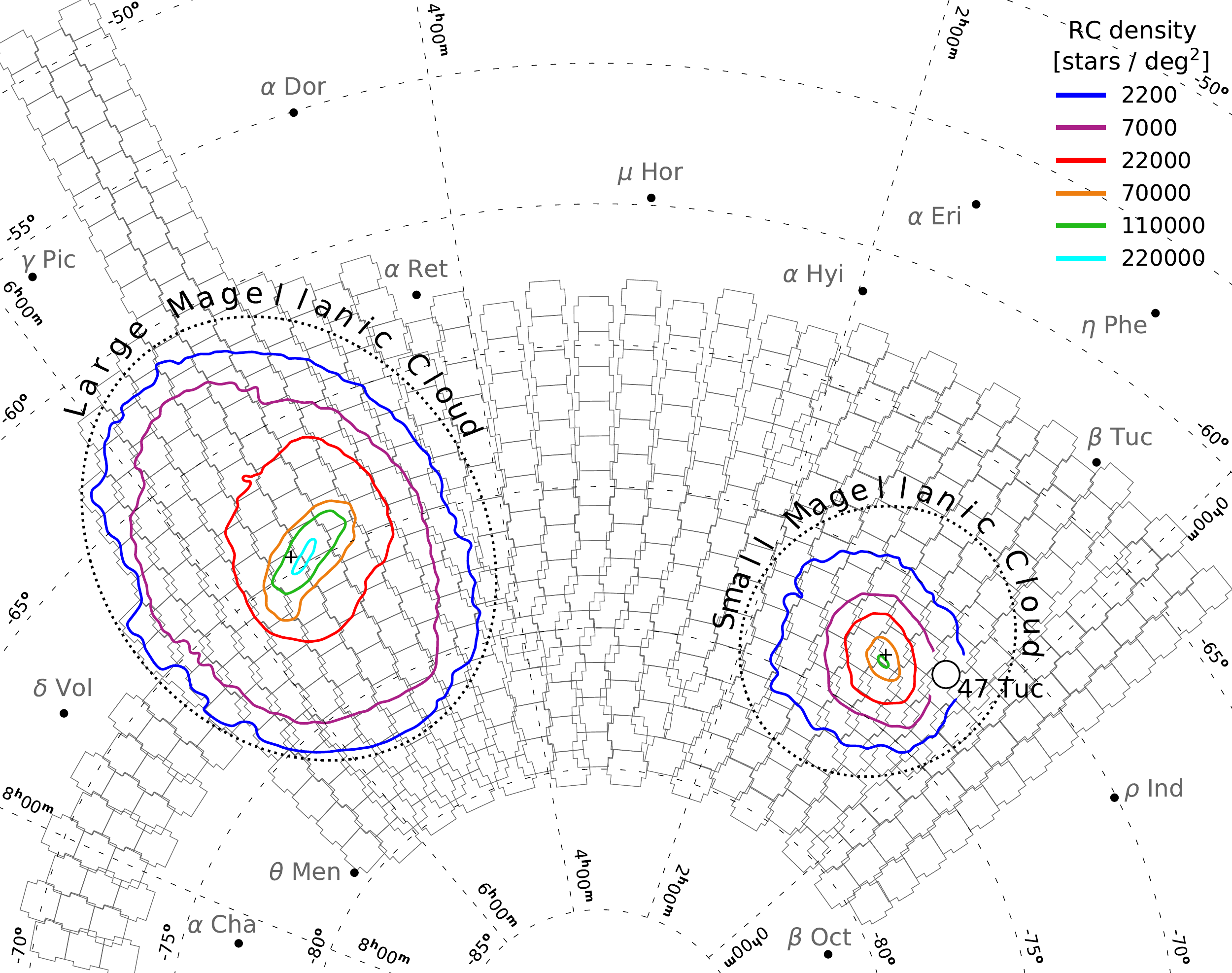}}
\caption{
OGLE-IV fields in the Magellanic Clouds region (approximately 670~deg$^2$
of the sky, 1.4~deg$^2$ camera field of view), marked with gray polygons.
Colored contours show densities of RC stars. The lowest density contour
(blue line) in the eastern part of the LMC is affected by the edge of the
OGLE footprint. Black crosses mark centers of the galaxies adopted
throughout this paper:  ($05^{\rm{h}}29^{\rm{m}}00^{\rm{s}}$, 
$-68^{\circ}30'00''$) from \cite{vanderMarel2001} for the LMC and
($00^{\rm{h}}52^{\rm{m}}12.5^{\rm{s}}$, $-72^{\circ}49'43''$) from 
\cite{deVaucouleurs1972} for the SMC.
}
\vspace{0.3cm}
\label{fig:fields}
\end{figure*}

Red Clump (RC) stars are low-mass red giants in the core He-burning stage of
the evolution that occupy a well defined region in the color-magnitude diagram
(CMD). This property allows for using these stars as standard candles for
determining distances \citep{Paczynski1998} and by measuring shifts between the
observed and theoretical RC color, can be successfully used to measure reddening
to these stars. For details on RC stars and their applications with possible
limitations see the review by \cite{Girardi2016}.

In this study we present the most detailed and extensive reddening maps of
the MCs to date, based on RC stars. Section \ref{sec:data} provides details on
observations and data preparation. In Section \ref{sec:rccolor} we describe
the process of determining RC color and in Section \ref{sec:vi0} we analyze the
intrinsic RC color distribution in the MCs. Section \ref{sec:maps} describes
the final reddening maps and compares them with previous reddening maps of
the MCs. We summarize the paper in Section \ref{sec:summary}.

\section{Observations and Data Preparation}
\label{sec:data}
\subsection{OGLE-IV Observations}

The 1.3~m telescope of the OGLE survey is located at the Las Campanas
Observatory in Chile and has been observing the southern sky since 1996 on a
nightly basis. The fourth phase of the project started in 2010, when a new 32
chip mosaic CCD camera with a 1.4 square degree total field of view was
installed on the telescope. Since then, OGLE-IV has been observing the
Magellanic System in the {\it I}- and {\it V}-band with a cadence of 1-4 days,
later reduced to about 10 days in the sparsely populated regions of the
Magellanic Bridge and the outskirts of the MCs.
The OGLE-IV fields covering approximately 670 square degrees in the Magellanic
System are pictured in Fig.~\ref{fig:fields}. For technical details on the
OGLE-IV survey please refer to \cite{Udalski2015}.

The magnitude range of regular OGLE-IV images is $12-21$~mag in the
{\it I}-band and $12.5-21.5$ mag in the {\it V}-band. In order to increase
the effective depth of the survey, deep OGLE-IV images were constructed
from 2 to 100 high quality frames (depending on the field), with a median of
8 images in the {\it V}-band and 88 in the {\it I}-band.
The resulting magnitude limit of the deep images is approximately $23$~mag
in the {\it I}-band and $23.5$ mag in the {\it V}-band.
Even though the use of the deep images is not necessary for this particular
study, because the mean magnitude of RC stars in the MCs is much brighter
than the regular survey limit, it is useful due to their cleanness -- stacking
multiple images removes practically all artifacts and spurious detections
from the final frame, which are hard to fully account for otherwise (see
discussion in \citealt{Skowron2014}).

\subsection{Data Preparation}

Observational data were reduced with a standard OGLE reduction pipeline and
photometry was carried out with an image subtraction pipeline as described in
\cite{Udalski2015}.
Each of the 32 CCD chips is 2048x4096 pixels, but the size of the photometric
map of that chip is slightly larger (about 2200x4496 pixels) due to small
telescope shifts between pointings. For the same reason, star counts on the
chip edges are lower than in the main parts of the chip. In order to ensure
high data completeness, we removed about 50 pixels on each side of the map in
the x-direction and about 100 pixels on each side in the y-direction. The cut
was determined individually for each field and chip by finding a deviation
from the median star counts across the frame.
After these cuts, many frames still had an overlapping region, so in the
next step the duplicate entries were removed from overlapping chips.

The Magellanic System is rich in globular and open clusters. We removed
all clusters listed in the catalog of the Magellanic System clusters
\citep{Bica2008} and in the OGLE Collection of of Star Clusters
\citep{Sitek2016,Sitek2017} from the final sample, using the mean
value of both dimensions as a cluster diameter.

Finally, we transformed the data from the equatorial coordinates $(\alpha,
\delta)$ to Cartesian $(x, y)$ coordinates using a Hammer equal-area projection
\citep{Snyder1993} centered at $\alpha_c=3.3$~h and $\delta_c=-70$~deg.
The data were then subdivided into square regions using six bin sizes:
$x, y = 0.032, 0.016, 0.008, 0.004, 0.002, 0.001$, corresponding to
areas of approximately $3.34, 0.83, 0.21, 0.052, 0.013, 0.003$ square
degrees, respectively. Varying subfield size is useful for increasing the
resolution in the central parts of the MCs and decreasing it in the outskirts,
where the star counts are low. In other words, we created six different maps
of the Magellanic System, where each map has a different resolution.
We then repeated the process after shifting centers of the square regions
by half the bin size in both $x$ and $y$ direction. As a result, number
of regions in each maps has increased fourfold, thus further improving
the final resolution.

\section{Determining the Red Clump Color}
\label{sec:rccolor}

\begin{figure}[t]
\centerline{\includegraphics[width=0.48\textwidth]{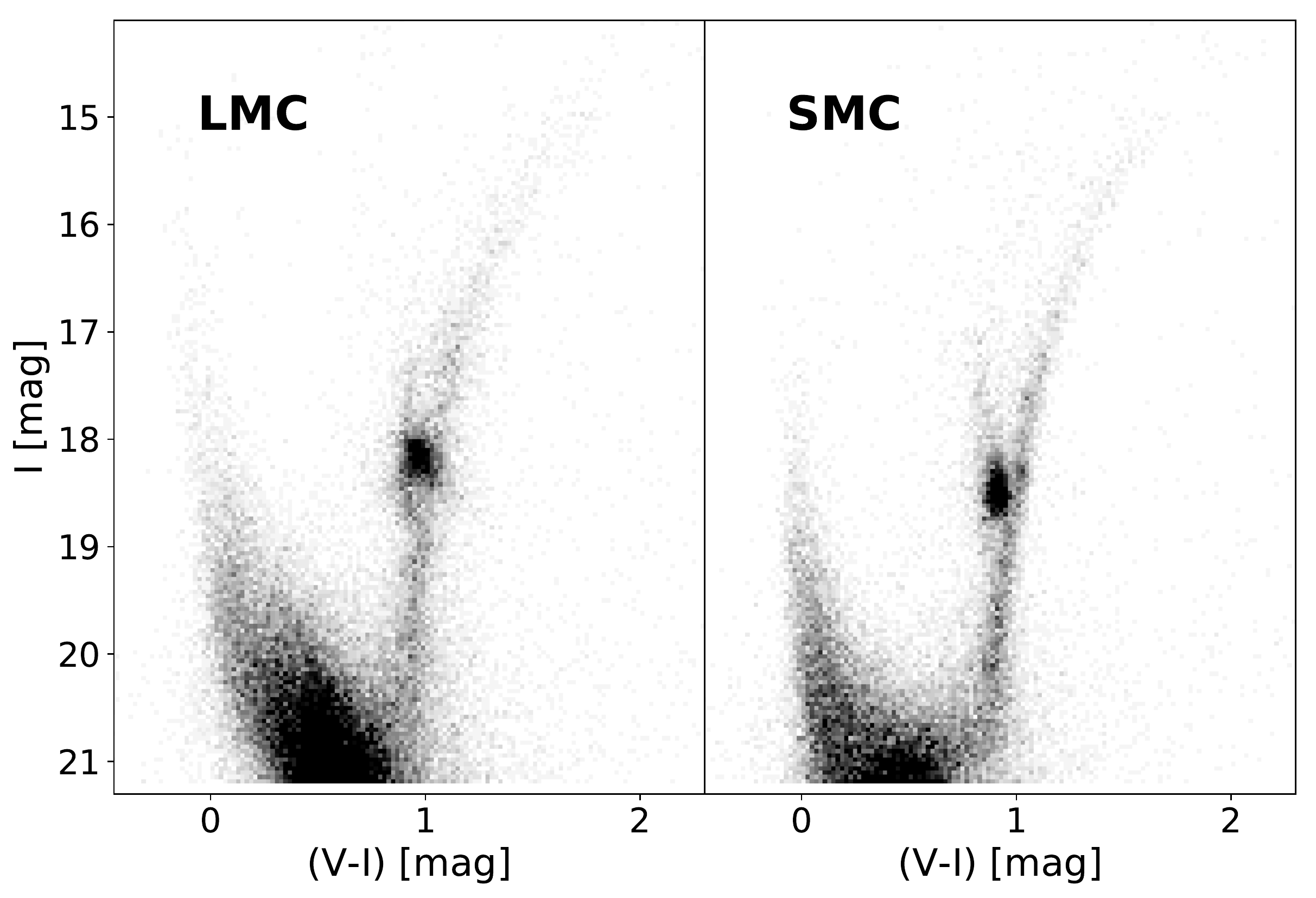}}
\caption{
Hess diagrams for a typical field in the LMC (left) and the SMC (right),
with a well visible RC and the RGB. The binning is 0.04~mag in magnitude
and 0.02~mag in color.
}
\vspace{0.3cm}
\label{fig:hess}
\end{figure}

For each subfield (and in each of the six subfield sizes), we construct a
$(V-I,I\,)$ color-magnitude diagram (CMD) and bin it in both magnitude and
color to obtain a Hess diagram, using bin sizes of 0.04 and 0.02 mag
in $I$ and $V-I$, respectively. The value of each bin is a number of stars
that fell into that bin. Fig.~\ref{fig:hess} shows an example of the Hess
diagram in the LMC (left) and in the SMC (right), with a well visible RC
and the red giant branch (RGB). The location of the RC on the CMD depends
on distance and metallicity and so it is different for the
two galaxies. The elongation of the RC in the SMC has been known for 
a long time \citep{Hatzidimitriou1989,Udalski2008b} and it has been shown
that it is caused by the large depth of the SMC along the line of sight,
rather than a presence of blue-loop stars \citep{Nidever2013}.

\vspace{0.1cm}

The first step in determining the RC color is to define the initial CMD region
occupied by RC stars, that will be used for a more detailed analysis. Its
location and extent changes with extinction -- the more reddened the region,
the more dispersed and shifted toward fainter magnitudes and redder colors the
RC is. There are two main approaches: 
a) to define a generous CMD region, which would cover all possible RC
locations, and deal with disentangling the RC in a subsequent algorithm,
b) to define a small box in $(V-I\,)$ and magnitude and modify its position and
size iteratively (or by hand) for each line of sight, and then use a simple
algorithm such as a mean or median to calculate the final RC color.
The disadvantage of the first approach is that the contamination from other
stars within the region forces the algorithm to be much more complex. However,
in the second approach, in order to remove the majority of contamination (e.g.,
the RGB) and outliers (from the MW foreground) one has to reduce the size of
the box and potentially remove some of the genuine RC stars. This leads to
biases in the final measurement.

Within the discussed scenarios the box can be
``rectangular'' or ``slanted'', i.e., the CMD region can be defined only by
specifying a range in color and magnitude, or it can follow the extinction
vector. In the second case, the area of the box can be safely reduced, which
in effect lowers the contamination from the RGB stars. We see no obvious
disadvantages of the latter choice, which can be simply viewed as a definition
of the magnitude axis using a Wesenheit index ($W$).

\begin{figure}[t]
\centerline{\includegraphics[width=0.5\textwidth]{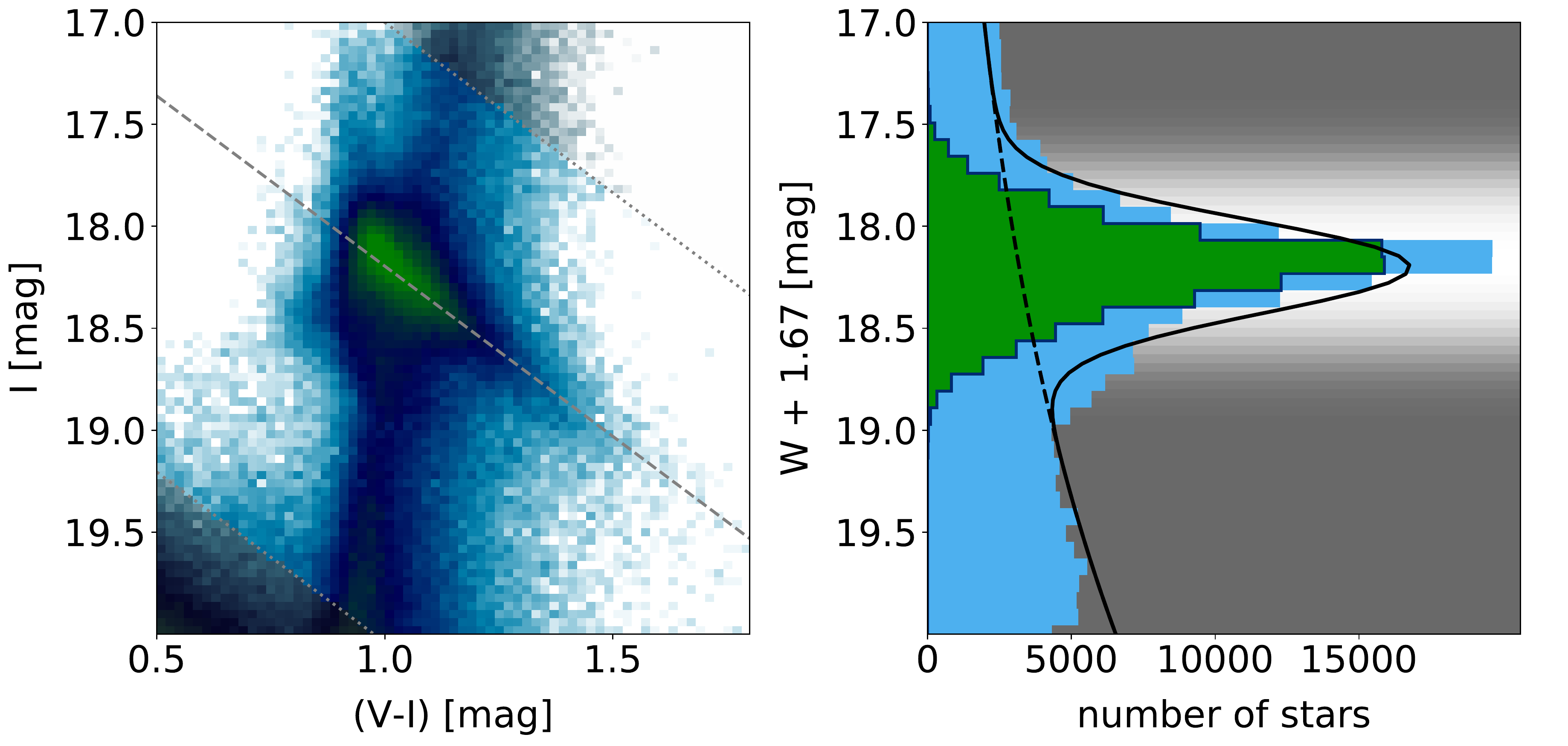}}
\caption{
Hess diagram for a typical moderately reddened field in the LMC (left panel)
with a well defined RC. Gray dotted lines mark the slanted CMD region that
is used for determining a luminosity function, and the dashed line shows the 
subsequently estimated peak of the RC distribution in $W$ 
(defined as $I - 1.67 (V-I\,)$). The magnitude histogram
from the entire slanted region (right panel) is shown in blue. The black
solid line is the fit to the histogram 
($\mathrm{LF} = \mathrm{LF}_\mathrm{RC} + \mathrm{LF}_\mathrm{RGB}$) 
while a dashed line is the RGB component ($\mathrm{LF}_\mathrm{RGB}$). 
The white, shaded band represents the probability 
($p_\mathrm{LF} = \mathrm{LF}_\mathrm{RC} / \mathrm{LF}$) of 
belonging to the RC, while the green histogram is that probability 
multiplied by the blue histogram, i.e. it can be interpreted as 
a magnitude distribution of RC stars. 
}
\vspace{0.3cm}
\label{fig:luminosity}
\end{figure}

In this study we choose the generous CMD region approach and the ``slanted''
magnitudes (left panel of Fig.~\ref{fig:luminosity}):
\begin{gather}
 W = I - 1.67 \; (V-I\,) \;\;\; \mathrm{for \;\; LMC}\\
 W = I - 1.74 \; (V-I\,) \;\;\; \mathrm{for \;\; SMC}
\end{gather}
We restrict our investigation to a conservative range in color of
$0.65 < (V-I\,) < 1.7$~mag and a magnitude range that falls within
$17 < I < 20$~mag at $(V-I\,) \approx 1$, which is typical for the RC.
In the center of the LMC, due to the large differential reddening we extend the
color range to $(V-I\,) = 2.1$~mag.
In our approach, the final results are not sensitive to the exact choice of
the extinction slope and magnitude range.

First, we determine a luminosity function of the RC in $W$, which is fairly
narrow (typical dispersion of 0.23 mag) even for highly extincted fields. 
There are two main components in the luminosity function: a distinct
distribution of stars belonging to the RC and a smooth contribution of the
RGB. The first is well approximated by the Gaussian  distribution and the
second can be modeled with an exponent or, in this narrow magnitude range,
even with a straight line (see the right panel of Fig.~\ref{fig:luminosity}).
Our aim is not to make a precise model of the luminosity function, but to
determine the most likely region in $W$ where the RC stars are located, and
then, with the measurement of the standard deviation, assign the probability
of belonging to the RC for each star.

We tested various fitting scenarios, in particular the normal distribution
plus the exponent, and found that the biggest obstacle in determining the RC
magnitude region is the unstable nature of the fits in regions with lower star
counts. Concurrently, we noticed that the scale parameter of the exponent is
rather consistent between various lines of sight. Hence, the exact value
of this parameter is not crucial, and we decide to fix it to $2.5$~mag, which
makes the fits more stable. The luminosity function then becomes:
\begin{equation}
\begin{aligned} 
\mathrm{LF}(W) &= \mathrm{LF}_\mathrm{RC}(W) + \mathrm{LF}_\mathrm{RGB}(W) \\
      &\sim exp\left( -\frac{1}{2}  \left(\frac{W - \overline{W}}{\sigma_W}\right)^2\right)  + A \; exp\left(\frac{W}{2.5}\right)
\end{aligned}
\end{equation}
where $\overline{W}$ is the mean, $\sigma_W$ is the standard deviation of the
Gaussian and $A$ reflects the relative amplitudes of the two components.
The luminosity function is marked with a black solid line in the right panel
of Fig.~\ref{fig:luminosity}.

\vspace{0.1cm}

\begin{figure}[t]
\centerline{\includegraphics[width=0.48\textwidth]{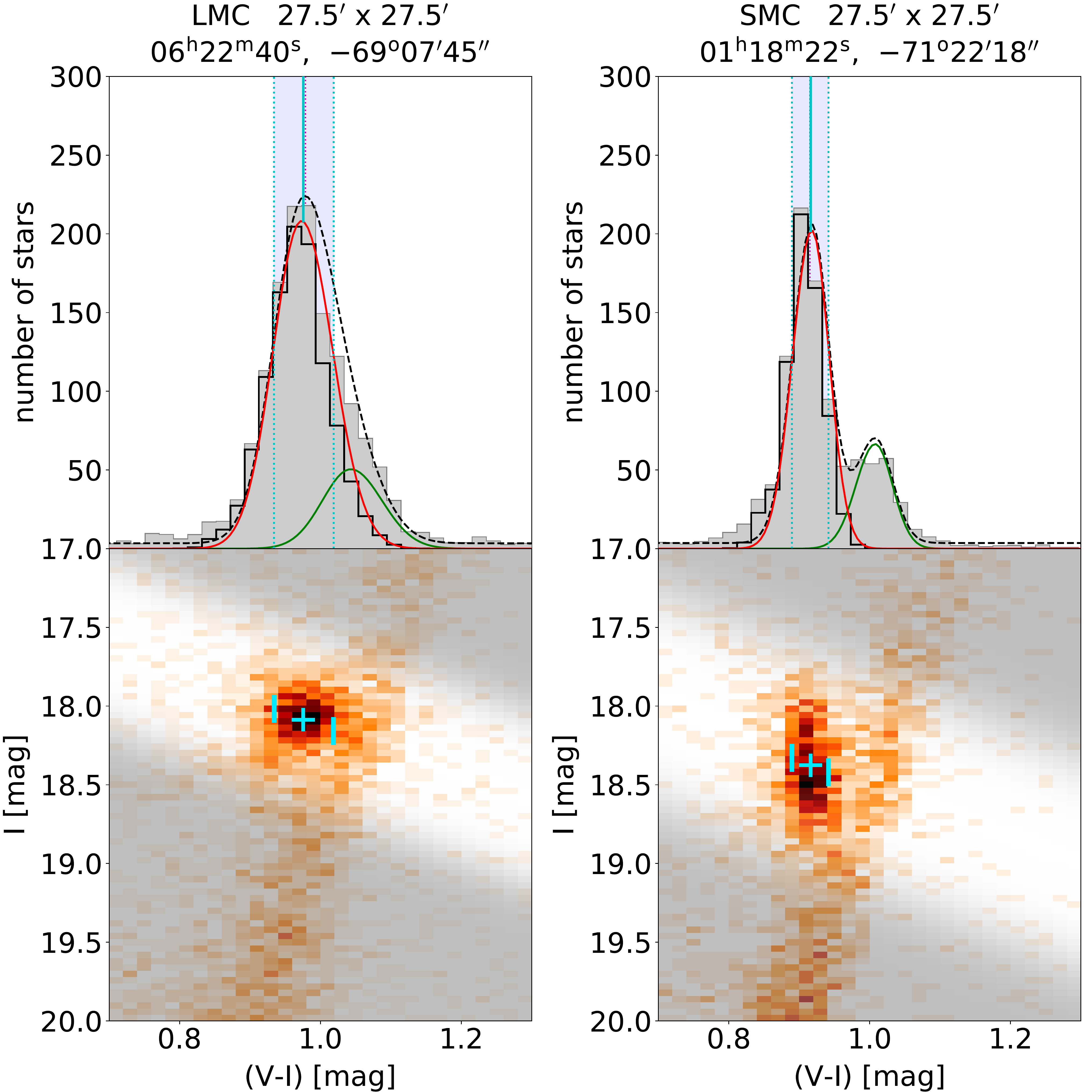}}
\caption{
Color distribution in the RC region for a typical field in the
LMC (left) and the SMC (right). Gray histograms in top panels show the
distribution of stars suspected to lie within the RC region based on the
luminosity function fit (histogram of all stars times the probability
$p_\mathrm{LF}(W)$). The dashed black line marks the whole model
$f(V-I\,)$ fitted to the distribution, while the red and green lines
show the RC and RGB components, respectively. The cyan vertical line is
the median of the RC component only and blue band mark 68\% interval.
Black histogram presents the final RC sample, where gray histogram is
multiplied by the probability based on the components of the fit
$p_\mathrm{RC}(V-I\,)$. Pink dashed line marks mode value of the RC color
sample. Bottom panels show Hess diagrams of the regions where the cyan
cross marks the median the fitted RC color distribution and cyan lines
correspond to the $\pm 34 \%$ intervals adopted from the upper panels.
}
\vspace{0.3cm}
\label{fig:colorfit}
\end{figure}

In the second step we fit the $(V-I\,)$ distribution where we weigh each star
with its probability of belonging to the RC sample based on the luminosity
function:
\begin{equation}
  p_\mathrm{LF}(W) = \mathrm{LF}_\mathrm{RC}(W) \, / \, \mathrm{LF}(W)
\end{equation}
This approach is analogous to choosing a box in magnitude $W$, but avoids
imposing sharp limits, which could bias the RC sample, or conversely, leave
too many unrelated stars within the sample. Even though this approach minimizes
the contamination of the RGB in the RC sample, it does not remove it completely
-- there is still a RGB component visible in the color distribution. The
majority of MCs regions have moderate differential reddening, and both
components can be well approximated using the normal distribution 
$\mathcal{N}(\,\overline{V-I}, \sigma)$ (see upper panels of
Fig.~\ref{fig:colorfit}).

\vspace{0.1cm}
 
\begin{figure}[t]
\centerline{\includegraphics[width=0.48\textwidth]{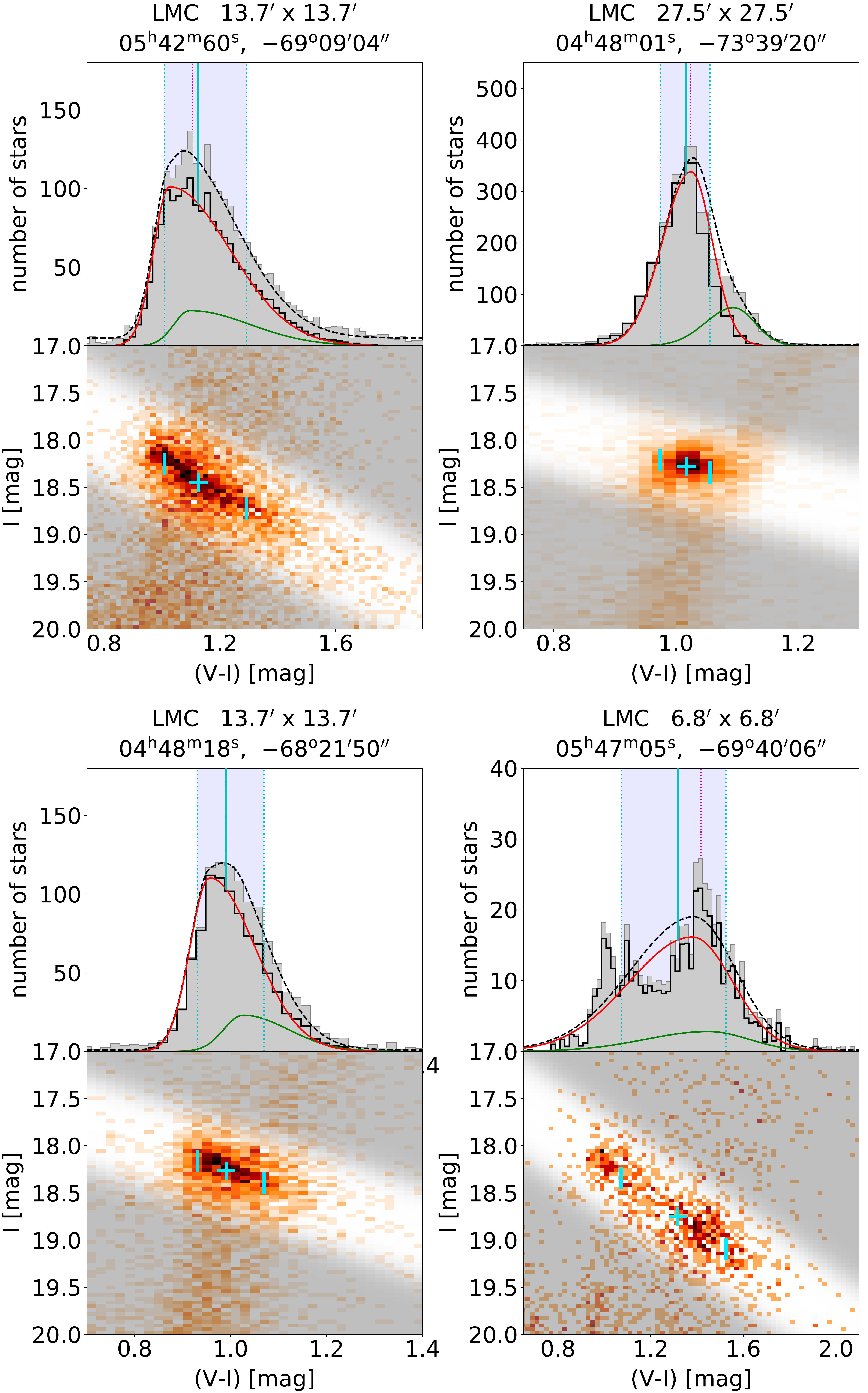}}
\caption{
Color distribution in the RC region for sample LMC fields with non-Gaussian
fits to the RC and RGB distributions. Lines and symbols are analogous to those
in Fig~\ref{fig:colorfit}. The combined distribution function ($f(V-I\,)$,
dashed gray) consist of a constant background term plus RC and RGB contributions
described by the $\mathcal{N}_2(V-I\,)$ model shifted by 0.07}
\vspace{0.3cm}
\label{fig:colorfit_ext}
\end{figure}

In order to properly model the RC color distribution in all lines of sight,
even those with a large differential reddening, we expand the basic model of
$\mathcal{N}(\,\overline{V-I}, \sigma)$ by allowing the $\sigma$ parameter to
be different for the low (L) and high (H) side of $\overline{V-I}$: 
\begin{equation}
\mathcal{N}_2(\,\overline{V-I}, \sigma_\mathrm{L}, \sigma_\mathrm{H}) = 
\begin{cases}
  \frac{ 2 \sigma_\mathrm{L}} {\sigma_\mathrm{L} + \sigma_\mathrm{H}} \mathcal{N}(\overline{V-I}, \sigma_\mathrm{L}), & \text{for } (V-I\,) < \overline{V-I} \\
  \frac{ 2 \sigma_\mathrm{H}} {\sigma_\mathrm{L} + \sigma_\mathrm{H}} \mathcal{N}(\overline{V-I}, \sigma_\mathrm{H}), & \text{for } (V-I\,) > \overline{V-I}
\end{cases}
\end{equation}
This modification was introduced after the visual inspection of hundreds of
fields with various models of the color fit. The most typical situation in the
inner parts of the MCs is that some fraction of stars lie between the observer
and the bulk of dust -- this fraction has a characteristic color scatter of
$\sigma_\mathrm{L}$. The rest of stars lie within or behind the dust and are
affected by the differential reddening, and for these stars, a better
approximation is to use a different, higher value of $\sigma_\mathrm{H}$.
Such a hybrid model also allows for a reasonable approximation of many atypical
RC distributions, but at the same time defaults to a Gaussian-like profile for
the majority of well-behaved lines of sight. Fig.~\ref{fig:colorfit_ext}
presents four examples of atypical RC color distributions, where this hybrid
model ($\mathcal{N}_2$) proves useful.

\vspace{0.1cm}

In general, Gaussian fitting is very sensitive to outliers which can introduce
a large bias in the fit parameters. Here we model the outliers with the constant
background component added to the components of the RC and the RGB. Our model in
principal has eight parameters: $(\overline{V-I})_\mathrm{RC}$,
 $(\overline{V-I})_\mathrm{RGB}$, four $\sigma$ components (L and H, for both RC
and RGB), $A_\mathrm{RGB}$, and $A_\mathrm{BKG}$, where the last two
parameters are the relative amplitudes of the RGB to RC and the background to
the RC, respectively.
Fitting of such model to the data is inherently unstable because the RGB and RC
regions overlap. Also the number of parameters is too large for many areas with
low star counts. But since we are interested in determining the color of the RC
only, and the RC component is dominant, we can determine other fit parameters
approximately. This is sufficient to reduce the impact of the RGB and the MW
foreground on the RC and reduces the systematic error of the RC color
measurement. In order to do so, we fix the color offset between the RGB and RC,
i.e. we fix $O_\mathrm{RGB, RC} = (\overline{V-I})_\mathrm{RGB} -
(\overline{V-I})_\mathrm{RC}$ to $0.07$ in the LMC and $0.09$ in the SMC.
These values were determined with a simple algorithm that estimates the RGB
color at the RC brightness using $\pm$ 1 magnitude range around the RC
(with the narrow $\pm\,2\,\sigma_W$ region around the RC magnitude removed). 
In other words, the algorithm uses parts of the RGB immediately above 
and below the RC. It then measures the median RGB color in 0.2 mag 
(or no less than 50 stars) bins, and using a linear fit, estimates the median RGB 
color at the RC position.
The reported color offsets between the median RC and RGB colors ($O_\mathrm{RGB, RC}$)
are practically constant across the galaxies, with a typical scatter of 0.01 mag.
In order to assess how such scatter influences the measured RC color, we
performed RC color calculations with $O_\mathrm{RGB, RC}$ changed by $-0.01$~mag and
$+0.01$~mag and found that it causes only a $-0.001$~mag and $+0.001$~mag
difference, respectively, in the calculated RC color. This shows, that assuming
a constant RGB-RC color offset in the entire galaxy is reasonable, especially
when other sources of error are an order of magnitude larger.

We also found that it is sufficient to use a single width of the Gaussian 
components for the RC and the RGB ($\sigma_\mathrm{L, RGB}=\sigma_\mathrm{L, RC}$
and $\sigma_\mathrm{H, RGB} = \sigma_\mathrm{H, RC}$) to model the observed
distribution, as a higher differential reddening affects both groups of stars in
the same way. This was verified by comparing standard deviations of the RC
and RGB distributions in all LMC and SMC sight lines. They are equal to within
a few percent.

Finally, we optimize the algorithm by estimating the relative number of stars
within the RGB and RC groups in the previous step, by integrating the RGB
luminosity function with appropriate weights ($p_\mathrm{LF}(W) \times
\mathrm{LF}_\mathrm{RGB}(W)$).

The four-parameter model:
\begin{equation}
\label{eq:modelf}
\begin{aligned}
f \big( (\overline{V-I})_\mathrm{RC}, &\sigma_\mathrm{L}, \sigma_\mathrm{H}, A_\mathrm{BKG} \big) \,(V-I\,) = \\
& f_\mathrm{RC} (V-I\,) +  f_\mathrm{RGB} (V-I\,) +  f_\mathrm{BKG} (V-I\,) \\ \\
f_\mathrm{RC} (V-I\,) =& \; \mathcal{N}_2 \big((\overline{V-I})_\mathrm{RC}, \sigma_\mathrm{L}, \sigma_\mathrm{H} \big)\,(V-I\,)  \\
f_\mathrm{RGB} (V-I\,) =& \; A_\mathrm{RGB} \, \mathcal{N}_2 \big((\overline{V-I})_\mathrm{RC} + O_\mathrm{RGB, RC}, \sigma_\mathrm{L}, \sigma_\mathrm{H} \big)\,(V-I\,)  \\
f_\mathrm{BKG} (V-I\,) =& \; A_\mathrm{BKG}
\end{aligned}
\end{equation}
(where $A_\mathrm{RGB}$ is calculated from the LF and $O_\mathrm{RGB, RC}$
is fixed) is fitted to the
color distribution of stars expected to lie within the RC region based on the
luminosity function fit, i.e., to the distribution of all stars multiplied by
the probability $p_\mathrm{LF}(W)$.

We calculate a median value (50th centile) of the first component of the fitted
model ($f_\mathrm{RC}$, representing only the RC distribution), as well as
$\pm 34.1\%$ intervals from the median, as an analog of a typical $\sigma$
value of a Gaussian distribution, providing the final color estimation as:
\begin{equation}
median(V-I\,)_{\;-\sigma_1}^{\;+\sigma_2} = median(V-I\,)_{centile(50 - 34.1) - median}^{centile(50 + 34.1) - median}
\end{equation}

The probability that a given star belongs to the RC sample can be assessed with:
\begin{equation}
  p_\mathrm{RC}(V-I\,) = f_\mathrm{RC}(V-I\,) / f(V-I\,)
\end{equation}
The black histograms in Figures~\ref{fig:colorfit} and~\ref{fig:colorfit_ext} 
show the final RC sample. The mode of the histogram (or the most likely color
value) is equal to the median for symmetric Gaussian profiles and is typically
smaller for asymmetric profiles, since for the majority of lines of sight, the
differential reddening makes $\sigma_\mathrm{H} > \sigma_\mathrm{L}$.
We estimate the modal value by fitting a constant level  with a small Gaussian
bump to the final RC sample, across the entire color range. The mean and sigma
of the Gaussian component are fitted, but the area under the Gaussian is fixed
and set to 1/6 of the total area of the model. That way, we avoid noise in the
mode which is typically imposed by arbitrary binning.
In the final table, we provide both the median and the mode of the RC color
distribution and leave it to the user to decide which is more suitable for
their needs. One can imagine that depending on the scientific question
asked, either the mode or the median reddening for a given line of sight
will give a better answer. However, in the subsequent analysis we will
use the median RC color and hence the median reddening value.
We also provide $\pm 34.1\%$ percentile values from
the median as a measure of confidence intervals and of differential reddening.

Previous studies of reddening in the Magellanic Clouds typically used a
combination of a Gaussian and a polynomial to fit the RC distribution
(e.g. \citealt{Haschke2011,Choi2018,Gorski2020}). However, this has a tendency
to miss the RGB contamination entirely and focus the polynomial on the smooth
stellar background  instead. In effect, in fields with low differential
reddening, this overestimates the RC color due to the RGB contamination, and
in fields with high differential reddening, it chooses the most likely value.
Additionally, in regions with lower star counts -- either in the outskirts of
the galaxies or in areas with large reddening, a simpler fit (with fewer
parameters) is generally more stable since it doesn't require a large number
of stars to converge. This allows for retaining high resolution in such areas.


\subsection{The Age-related Complexity of the Red Clump}
\label{sec:complexity}

\begin{figure}[t]
\centerline{\includegraphics[width=0.46\textwidth]{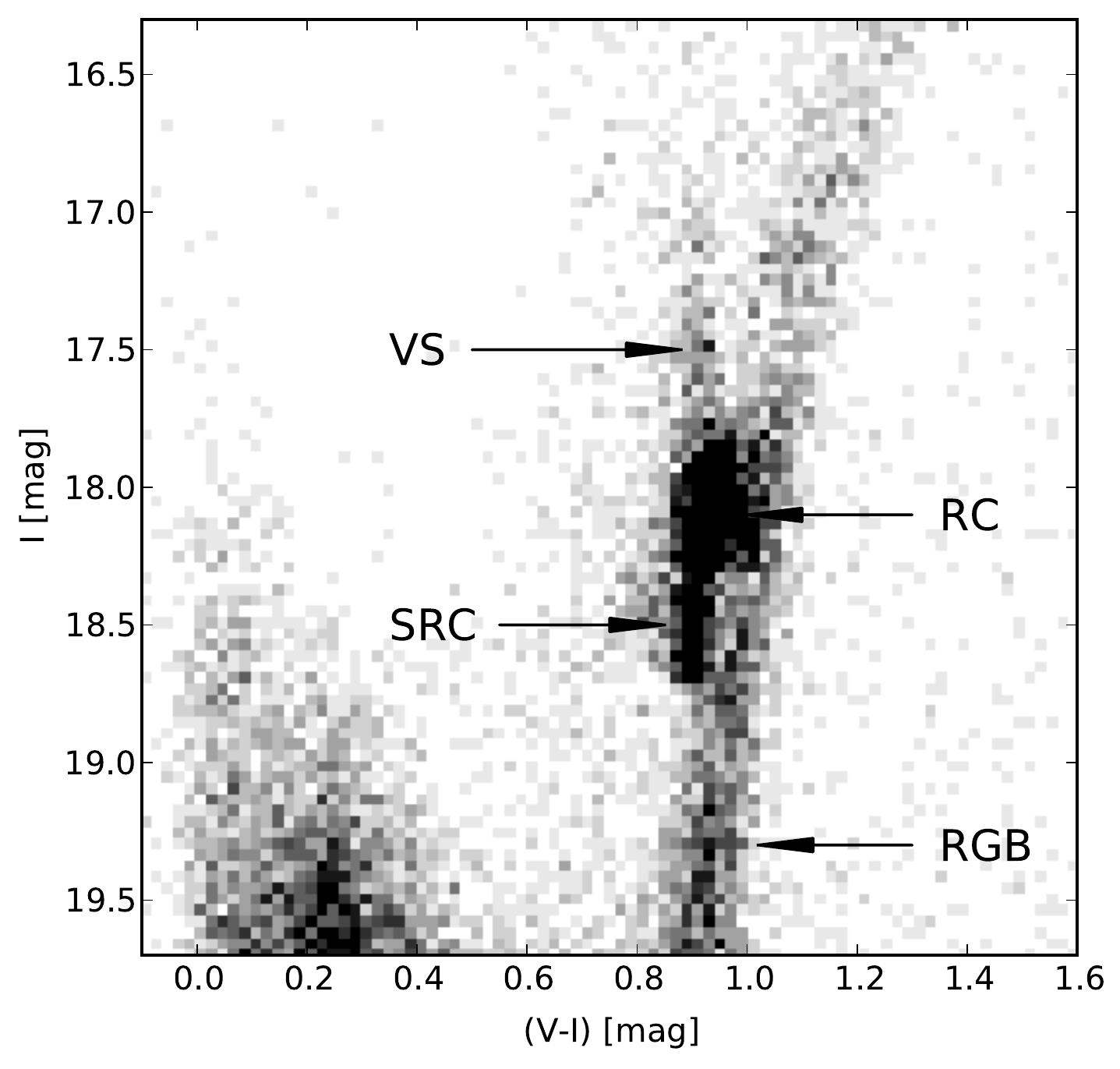}}
\caption{
A Hess diagram of a 27.5$'$ x 27.5$'$ area in the LMC (centered at
$\alpha= \,$05:30:52, $\delta=-$63:49:12) showing an example of the complexity
of the RC. The marked structures are the main RC, the Secondary RC (SRC) and
the Vertical Structure (VS).
}
\vspace{0.3cm}
\label{fig:rc_structures}
\end{figure}

The RC observed in some parts of the LMC is rich in various substructures:
a distinct secondary RC (SRC), which is about $0.4$~mag fainter and slightly
bluer than the main RC and is composed of younger, $\sim 1$~Gyr stars
\citep{Girardi1999}; a vertical structure (VS), which is made of brighter
and more massive RC stars; a horizontal branch (HB) composed of low-mass
metal-poor stars at the same evolutionary stage (see the review by
\citealt{Girardi2016}). Fig.~\ref{fig:rc_structures} shows an example of these
substructures in one of many such sight lines in the LMC.
The SRC is the second most pronounced feature in the CMD and its origin is well
pictured in Fig.~11 of \cite{Girardi2016} which shows the age structure of RC
stars -- the younger the stars within the $1-2$~Gyr range, the bluer they are.
Also, as shown in Fig.~6 of \cite{Girardi2016}, their luminosity varies by
$\sim 0.5$~mag within this age range, which is the reason why the SRC is
elongated in magnitude. However, for stars older than $2-3$~Gyr, both magnitude
and color change very slightly with age (see Fig.~1 of \citealt{Girardi2001}),
resulting in a very compact RC. In other words, the $(V-I\,)$ of older RC stars
remains almost constant with age.

\vspace{0.1cm}

\begin{figure}[t]
\centerline{\includegraphics[width=0.48\textwidth]{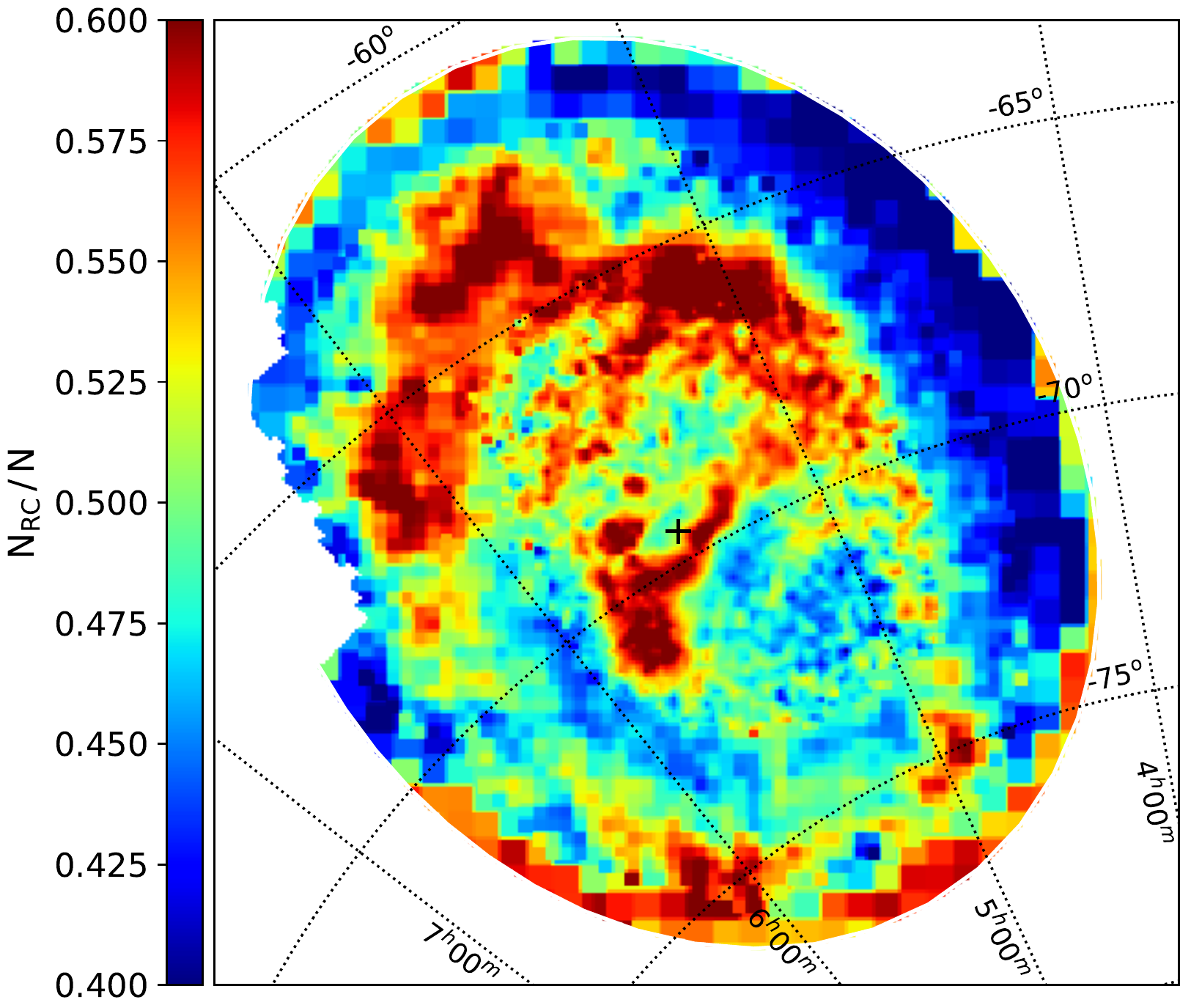}}
\caption{
The two-dimensional distribution of ${\rm N_{RC} / N}$ -- the
relative number of RC stars to all stars in the $\pm\,1$~mag range around
the measured RC brightness, in the LMC. The detailed discussion of the
observed distribution will be presented in Skowron et al. (2021,
{\em in preparation}).
}
\vspace{0.3cm}
\label{fig:dino}
\end{figure}

The presence of the SRC and/or VS significantly widens the observed distribution
of RC stars in magnitude ($\sigma_W$) and influences our estimation of the
number of stars in the RC (${\rm N_{RC}}$) based on the LF. 
In order to trace regions of the LMC where these additional structures are
visible, we calculate the  ${\rm N_{RC} / N}$ parameter, which is the
relative number of RC stars to all stars in the $\pm\,1$~mag range around the
measured RC brightness ($W_\mathrm{RC}$). In other words, it is the ratio
of the number of stars in the green histogram to the number of stars in the blue  
histogram shown in the right panel of Fig.~\ref{fig:luminosity}, but limited 
to the $\pm\,1$~mag range around $W_\mathrm{RC}$.  In regions where this
parameter is above $\sim 50\%$, we observe the additional RC structures which
are significant enough, that may influence the color measurement of the older,
main RC. The two-dimensional distribution of ${\rm N_{RC} / N}$ is shown
in Fig.~\ref{fig:dino} and will be discussed in detail in a forthcoming paper
(Skowron et al. 2021, {\em in preparation}).

\vspace{0.1cm}

The presence of an additional young ($< 2$~Gyr) RC population, whose
intrinsic color is bluer by a couple of hundredths of a magnitude
($\sim 0.04$~mag) than that of a regular RC, has an effect on the measurement
of $(V-I\,)$. The resulting offset is of the order of $0.01$~mag and depends
on the relative number of the young to old RC population. Since the regular RC
is more numerous and its intrinsic color is more stable across the galaxy, we
decide to use only the regular, older RC to derive the reddening.
The detailed modelling of a complex RC structure, including a star formation
history, would be a very challenging task itself, and is beyond the scope of
this paper. However, the influence of the young RC on the measured color of
the main RC can be estimated empirically using the ${\rm N_{RC} / N}$
parameter as a proxy for the severity of the effect, and this is done in
Section \ref{sec:vi0}.

In the SMC, the RC is largely elongated in magnitude. Some of this dispersion
may be explained by the RC age and metallicity distribution \citep{Girardi1999},
although the majority of this effect is caused by a large depth of the SMC along
the line of sight, rather than population effects \citep{Nidever2013}. Hence,
even though the distribution of ${\rm N_{RC} / N}$ is not entirely uniform
throughout this galaxy, this is caused mainly by geometrical, not population
effects and does not need to be taken into account.

\subsection{Discussion of $(V-\,I)$ uncertainties}

The uncertainties for a single OGLE photometric measurement are small
(about $0.005$~mag in the {\it I}-band and $0.008$ in the {\it V}-band
at the RC magnitude) and since those measurements are averaged
for at least 80 stars to obtain the RC color measurement for each
line of sight, their individual errors have a negligible effect
($\lesssim 0.001$~mag).

The accuracy of our final $(V-I\,)$ color of the RC is a combination of
three factors. The field-to-field systematic calibration errors of the OGLE
photometry are of the order of $0.01$~mag \citep{Udalski2015}. The second
source of error is associated with fitting the RC distribution and estimating
the median.
In order to assess the error of calculating the median color we performed a
number of simulations, where $50000$ star samples were generated from the
model $f$ (Eq.~\ref{eq:modelf}), with a varying number of stars and a varying
differential reddening parameter.
We then processed these artificial samples in the exact same way as the real
data and estimated the median color of the RC. The resulting error was from
about $0.001$~mag for samples containing $500-1000$ stars and with a small
spread in RC color ($\sigma \sim 0.02-0.04$~mag), to about $0.008$~mag for
samples containing only $80-200$ stars and having a large spread in RC color
($\sigma \sim 0.07 - 0.09$~mag). The typical error of the median RC color
across the galaxies (averaged for all lines of sight) is $0.0035$~mag and
$0.0031$~mag, in the LMC and SMC, respectively.

The third source of uncertainty in estimating $(V-I\,)$
is the difference between the assumed model of the color distribution
and the actual distribution of stars -- the intrinsic shape of the RC and RGB
populations does not necessarily follow a simple Gaussian profile, but it
changes with the star-formation history and metallicity distribution.
However, the dominant difference comes from the complex nature of dust
clouds in any given field and the relative position of stars and dust clouds
along the line of sight.
Our $\sigma_1$ and $\sigma_2$ parameters are indicative of the variability of
the reddening in a given field. These are also our best estimations of the
uncertainty of the reddening value toward a given star.
In the majority of the MCs $\sigma_1$ and $\sigma_2$ are typically
between $0.03$ and $0.05$ mag. In the centers they are of the order of
$0.10$~mag, reaching $0.30$~mag in the most dusty regions of the LMC.

In summary, the errors of estimating the median $(V-\,I)$ in regions with low
$\sigma$ values are not greater than the systematic uncertainties of the
photometry, while with the growing values of $\sigma$ the uncertainty
of estimating the median becomes less and less relevant.
Nevertheless, we can provide an upper bound on the error of our estimations
for the majority of the studied area. In Sections \ref{sec:vi0_lmc} and
\ref{sec:vi0_smc} we measure the scatter between our median reddening
values and values interpolated from the reddening maps of \cite{Schlegel1998}
in the outer parts of both galaxies. We find the scatter of $0.014$~mag in the
LMC and $0.018$~mag in the SMC in all lines of sight outside of the central
regions. These values encapsulate all sources of error from both our and
\cite{Schlegel1998} study, as well as other systematic problems, such as:
the mismatch of the spatial resolution between the two studies; the assumptions
about the intrinsic color of the RC;  and the reddening law used to transform
reddening values between different passbands.

\section{Determining $(V-I\,)_0$ of the Red Clump}
\label{sec:vi0}

The intrinsic color of the RC depends on both age and metallicity
\citep{Girardi2001}. Widely used reddening maps of \cite{Haschke2011} adopted
a theoretical value of 0.92~mag and 0.89~mag for $(V-I\,)_0$ in the center of
the LMC and SMC, respectively. \cite{Gorski2020} measured $(V-I\,)_0=0.838$~mag
for the LMC and $(V-I\,)_0=0.814$~mag for the SMC, which are a mean of four
distinct methods. \cite{Nataf2020} employed a hybrid theoretical and empirical
approach and derived $(V-I\,)_0 = \{0.89, 0.92, 0.88\}$~mag for the LMC, using
three different methods, and $(V-I\,)_0 = \{0.84, 0.84\}$~mag for the SMC.
All of the above studies were limited to the inner parts of the MCs and 
assumed a constant metallicity, and thus the RC color value.

In order to determine the intrinsic unreddened $(V-I\,)_0$ color of the RC in
the MCs, we correct the median measured $(V-I\,)$ values for the
foreground dust of the Milky Way (MW), for which we use the all-sky reddening
map of \cite{Schlegel1998} based on dust infrared emission (hereafter SFD). The
map provides $E(B-V)$ values that can be converted to $E(V-I\,)$ taking into
account the recalibration of the SFD dust map by \cite{Schlafly2011}. The
recalibration recommends a 14\% decrease of $E(B-V)$ and the use of
$R_V = 3.1$ \cite{Fitzpatrick1999} reddening law. We use the convenient
conversion coefficients provided by \cite{Schlafly2011}:
\begin{gather}
A_V/E(B-V\,)_\mathrm{SFD,original} = 2.742 \\
A_I/E(B-V\,)_\mathrm{SFD,original} = 1.505
\end{gather}
that give 
\begin{equation}
E(V-I\,)_\mathrm{SFD} = A_V-A_I = 1.237 \; E(B-V\,)_\mathrm{SFD,original}.
\end{equation}

\begin{figure}[t]
\centerline{\includegraphics[width=0.5\textwidth]{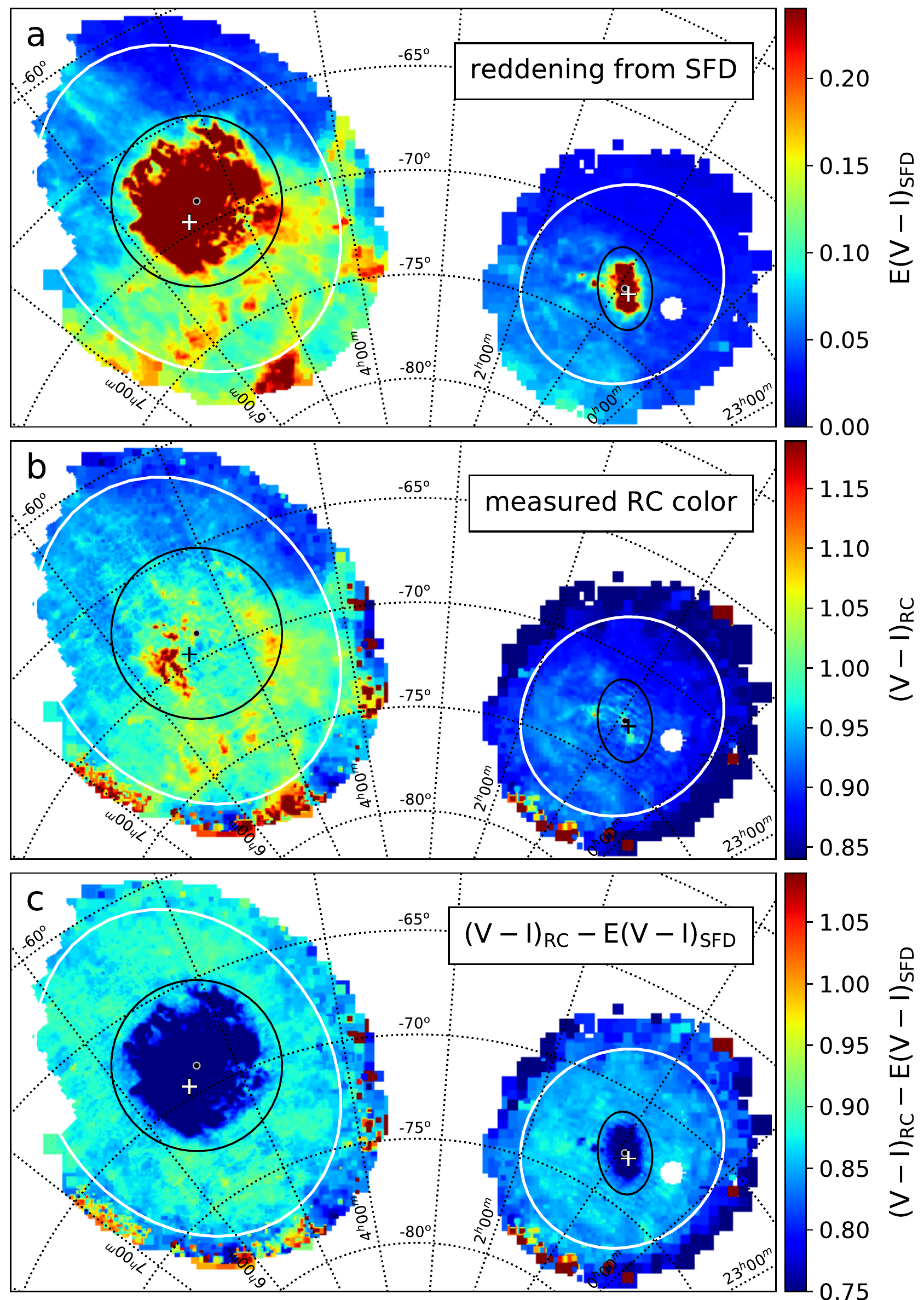}}
\caption{
The comparison of SFD reddening in the MCs area with RC color measured
in this work. Panel~{\bf a} shows the Galactic reddening map from SFD 
converted to $(V-I\,)$. Panel~{\bf b} presents RC color values measured
in this work (median), while panel~{\bf c} shows the difference
between panels {\bf b} and {\bf a}, i.e. the RC colors corrected for
the MW reddening. Central regions of the MCs, in which the SFD map is
not reliable, are marked with black circle and ellipse for the LMC and SMC,
respectively, while the rejected outer regions are marked with white
ellipses. Black dots represent dust disk centers of the MCs, and the
crosses mark the optical centers of the LMC and SMC. The empty region
within the SMC is the location of the globular cluster 47~Tuc.
}
\vspace{0.3cm}
\label{fig:sfd}
\end{figure}

\begin{figure}[t]
\centerline{\includegraphics[width=0.5\textwidth]{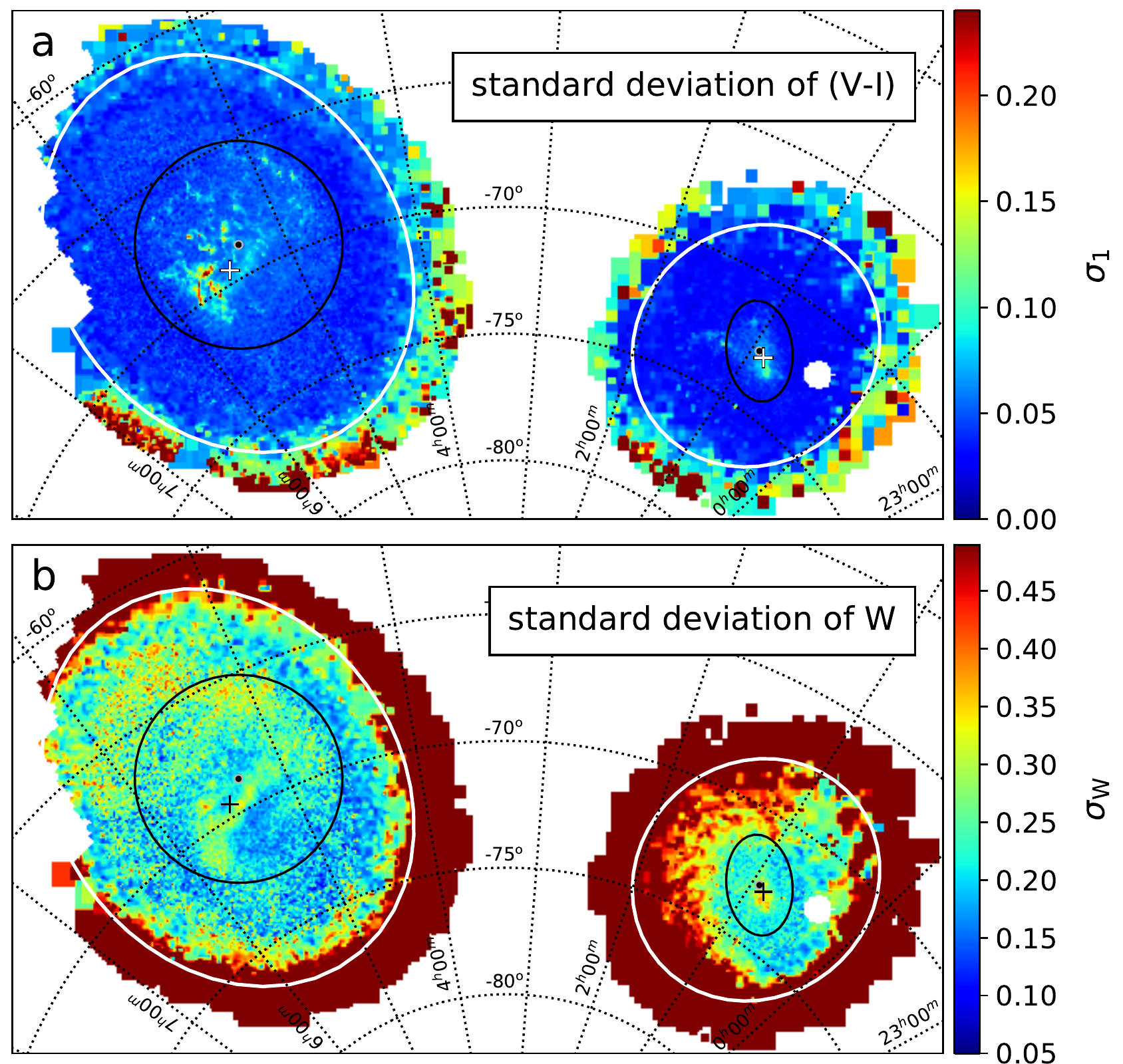}}
\caption{
The scatter of RC color $\sigma_1$ (panel~{\bf a}) and the standard
deviation of RC brightness $\sigma_W$ (panel~{\bf b}).
Central regions of the MCs, in which the SFD map cannot be used, are
marked with black circle and ellipse for the LMC and SMC, respectively,
while the rejected outer regions are marked with white ellipses.
Black dots represent dust disk centers of the MCs, and the crosses mark
the optical centers of the LMC and SMC.}
\vspace{0.3cm}
\label{fig:sigmas}
\end{figure}

The SFD galactic reddening map is shown in panel~{\bf a} of Fig.~\ref{fig:sfd}.
The dust temperature and reddening values in the LMC and SMC centers are not
reliable since their temperature structure is not sufficiently resolved
\citep{Schlegel1998}, therefore we cut out the central regions of the MCs from
further analysis of $(V-I\,)_0$.
In the case of the LMC, we remove the inner $4.1^{\circ}$ radius circle
centered at $(05^{\rm{h}}20^{\rm{m}}00^{\rm{s}}$, $-68^{\circ}48'00'')$.
This region contains the inner disk of the LMC, which is abundant in gas, dust
and young stars. The dust content is clearly seen in SFD emission maps
(panel~{\bf a} of Fig.~\ref{fig:sfd}) and will be apparent in our maps of
differential reddening.
The majority of Classical Cepheids in the LMC \citep{Jacyszyn2016} are located
in this region (with a sharp decline in density outside).
This is also the center of rotation of HI gas measured using 21-cm line
\citep{deVaucouleurs1960}, and was also shown to be an acceptable center for
the optical rotation curve \citep{Feast1961}.
Thus we adopt this ``radio center'' coordinates as a center of the excluded
region, which is offset by over one degree from the LMC optical disk center
at ($05^{\rm{h}}29^{\rm{m}}00^{\rm{s}}$, $-68^{\circ}30'00''$) from
\cite{vanderMarel2001}.

In the case of the SMC, we remove measurements within an ellipse centered at
$(00^{\rm{h}}58^{\rm{m}}00^{\rm{s}}$, $-72^{\circ}12'00'')$, that was selected
by eye to encompass the affected central region of the galaxy
($r_a=2.0^{\circ}$, $r_b=1.3^{\circ}$, $r_{PA}=40^{\circ}$)
and is also different from the optical center of the SMC at
($00^{\rm{h}}52^{\rm{m}}12.5^{\rm{s}}$, $-72^{\circ}49'43''$) from
\cite{deVaucouleurs1972}.
See black circle/ellipse in Fig.~\ref{fig:sfd}, with their centers marked with
black dots and optical centers marked with crosses.

\vspace{0.1cm}
\begin{figure}[t]
\centerline{\includegraphics[width=0.48\textwidth]{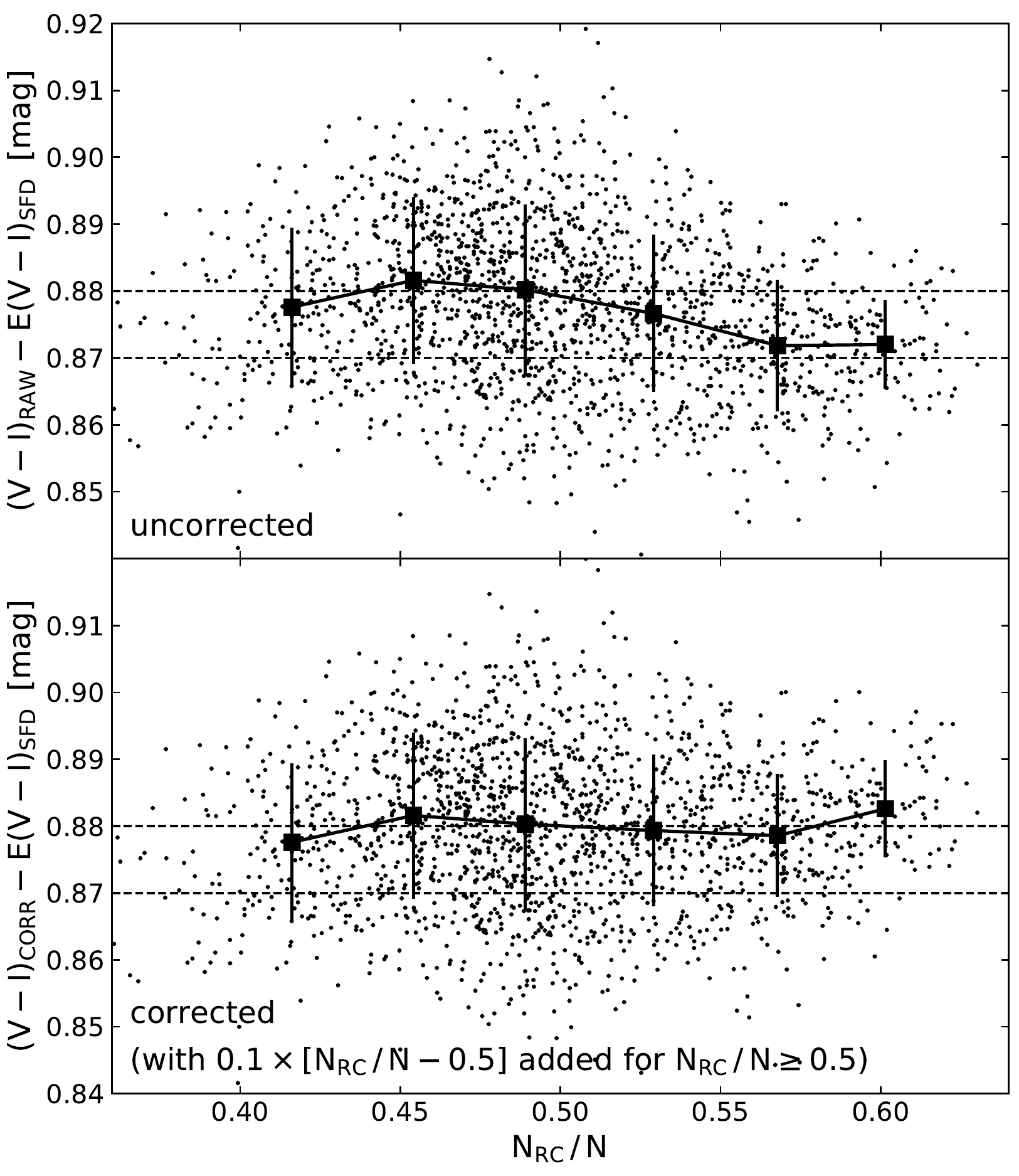}}
\caption{
The dependence of the measured RC color on ${\rm N_{RC} / N}$.
The top panel shows the change of our measured intrinsic color $(V-I\,)_0=
(V-I\,)_{\rm RC} - E(V-I\,)_{\rm SFD}$ with the growing contribution of
the SRC to the total RC. The mean change is not more than 0.01~mag.
The bottom panel shows the above corrected for the presence of the SRC,
according to the simple straight line fit as written in the panel.
}
\vspace{0.3cm}
\label{fig:dino_offset}
\end{figure}

The middle panel of Fig.~\ref{fig:sfd} shows the median $(V-I\,)$ colors
of the RC in the MCs measured in this work, while panel~{\bf c} shows the RC
colors corrected for the MW reddening (the difference between panels {\bf a}
and {\bf b}), i.e. the intrinsic color $(V-I\,)_0$.
Once again, it is clear that SFD reddening values within the central regions
of the MCs are not real and only the values outside the marked areas can be
used to determine the intrinsic color of the RC.
Panels {\bf b} and {\bf c} also illustrate, that the outskirts of the galaxies
(the areas outside white ellipses) are not suitable for determining $(V-I\,)_0$
-- the color determination in these regions is erratic and the visual
investigation of color fits in these lines of sight confirms that they are not
reliable. The main reason is that we reach the edge of the galaxies where the
RC star counts are so low, that the RC is not well defined. This is well
pictured in panel~{\bf a} of Fig.~\ref{fig:sigmas}, where we see a significant
increase in $\sigma_1$ in the outskirts of the MCs (panel  {\bf a}).
The highest values of $\sigma_1$ coincide with the increased MW
extinction in the southern part of the LMC. This is caused by the reddened
population of MW stars interfering with the RC of the LMC in the CMD.
As a result, the region occupied by RC stars in the CMD is dominated by the MW
foreground and the Gaussian fit generates a very wide peak across the entire
fitting box defined in Section~\ref{sec:rccolor}. This is also reflected in
a very high value of $\sigma_{W}$ (panel~{\bf b}), where the outskirts of the
MCs reach $\sigma_{W}=0.5$~mag,
which reflects the fact that the LF is dominated by the background contamination
and there is no distinct RC component visible.

We therefore assume that all measurements that fall outside the white ellipses
(Figs.~\ref{fig:sfd}-\ref{fig:sigmas}) are unreliable and reject them from
the intrinsic RC color analysis. The ellipses were chosen based on the surface
density of RC stars -- about 1000~stars/deg$^2$.
Then, after visual inspection of CMDs on both sides of the ellipses,
the ellipse parameters were slightly adjusted to ensure that the bad
measurements are outside the ellipse, while the majority of the good ones
fall inside the ellipse.

As discussed in Section~\ref{sec:complexity}, there is a bias in the measurement
of $(V-I\,)$ of the RC in some parts of the LMC that is caused by the presence
of additional RC structures, mainly the younger SRC. The additional, bluer SRC
is composed of stars younger than about 2~Gyr, while the majority of RC stars
are older, and their intrinsic color $(V-I\,)_0$ remains fairly constant with
age. Fig.~\ref{fig:dino} shows the distribution of ${\rm N_{RC} / N}$ that
reveals sight lines, where the measured $(V-I\,)$ is affected by the young SRC. 
Fig.~\ref{fig:dino_offset} shows how our measurement of the intrinsic color is
in result biased with the growing value of ${\rm N_{RC} / N}$.
Even though the color difference between the regular RC and the SRC is a few
hundredths of a magnitude, typically $\sim 0.04$~mag, its influence on the
measured color is on average only of the order of 0.01~mag. We apply this small
correction ($0.1 \times ({\rm N_{RC} / N} - 0.5)$ for
${\rm N_{RC} / N} \geq 0.5$) to the final value of $(V-I\,)$, based on
Fig.~\ref{fig:dino_offset}.
The detailed modelling of the RC in individual sight lines is not necessary,
because the correction is small compared to other sources of uncertainty.

\subsection{Metallicity gradient in the MCs}
\label{sec:feh_gradient}

As was already discussed, the RC color depends on both age and metallicity
of the RC. Since we are using only the intermediate and old age RC stars
(approximately $> 2-3$~Gyr), the color change with age is negligible in
this age range, but the change with metallicity is not (Fig.~1 of
\citealt{Girardi2001}).

The presence of the metallicity gradient in the LMC has long been a subject
of discussion (\citealt{Choudhury2016} and references therein), although
recent studies support a shallow metallicity gradient, e.g.:
$-0.047 \pm 0.003$ dex/kpc out to $\sim 8$ kpc \citep{Cioni2009},
$-0.029 \pm 0.002$ dex/kpc in the inner 5~kpc \citep{Skowron2016}, and
from $-0.049 \pm 0.002$ to $-0.066 \pm 0.006$ dex/kpc up to a radius
of 4~kpc \citep{Choudhury2016}.
The distribution of [Fe/H] among the LMC clusters also shows a radial
dependence \citep{Pieres2016}.

In the case of the SMC, the existence of the metallicity gradient has
also been questioned, although a number of most recent results do report
a shallow metallicity gradient:
$-0.075 \pm 0.011$~ dex/deg within the inner $5^\circ$ \citep{Dobbie2014},
$-0.08 \pm 0.08$~ dex/deg out to $4^\circ$ \citep{Parisi2016},
from $-0.045 \pm 0.004$ to $-0.067 \pm 0.006$ dex/deg within the radius of
$2.5^\circ$ \citep{Choudhury2018}, and
$-0.031 \pm 0.005$ within the inner $2^\circ$ \citep{Choudhury2020}.

\begin{figure}[t]
\centerline{\includegraphics[width=0.48\textwidth]{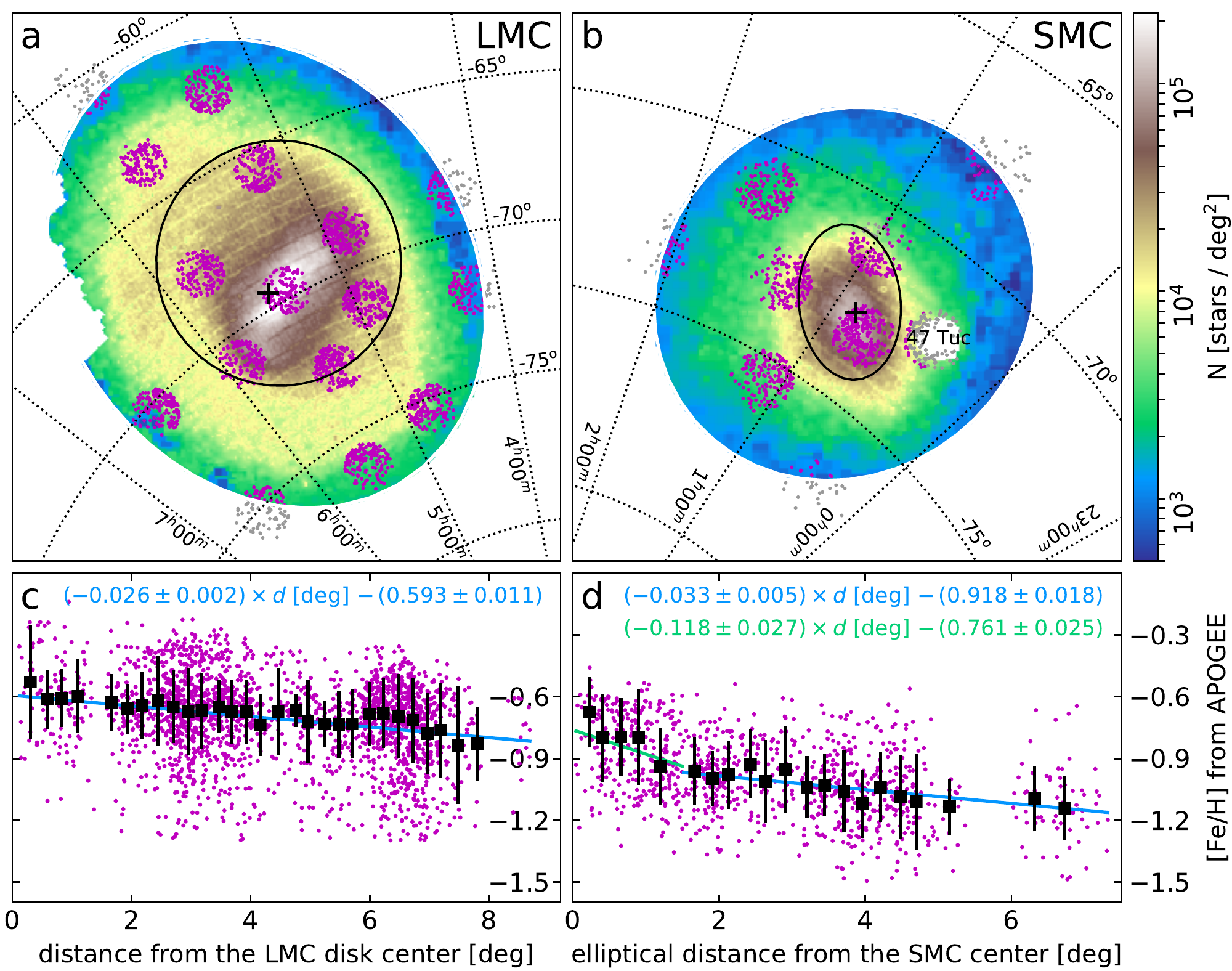}}
\caption{
The sample of red giants from APOGEE MCs fields \citep{Nidever2020}.
Panels {\bf a} and {\bf b} show their distribution on top of the
density maps of OGLE-IV giants, for the LMC and SMC, respectively.
Panels {\bf c} and {\bf d} show how spectroscopically determined
[Fe/H] values change with the distance from the center of the LMC
and SMC, respectively. The blue line in the bottom panels is the
line fit to the data. Black squares mark the binned data.
}
\vspace{0.3cm}
\label{fig:apogee}
\end{figure}

\cite{Nataf2020} used [Fe/H] metallicities of bright and faint red giants,
derived by the ASPCAP pipeline \citep{Garcia2016} from high resolution
spectra taken for the APOGEE survey \citep{Majewski2017}, which is part of
the Sloan Digital Sky Survey \citep{Blanton2017}, and found a mean
metallicity value for the inner LMC ${\rm [Fe/H]}=-0.64$~dex (within the
inner LMC disk, i.e. inside the black circle), and ${\rm [Fe/H]}=-0.88$~dex
for the inner SMC.
Following \cite{Nataf2020}, we select bright and faint red giants 
(${\rm K} > 12.25$~mag and $0.55 < {\rm J-K} < 1.3$~mag) as defined
by \cite{Nidever2020}, from APOGEE LMC fields 1-17, encompassing both the
central parts and the outskirts of the LMC, and from SMC APOGEE fields 1-7
(see their Fig.~1). We then remove foreground stars (${\rm log}(g) \geq 2$)
and very metal-poor stars (${\rm [Fe/H]} \leq -1.3$~dex in the LMC and
${\rm [Fe/H]} \leq -1.6$~dex in the SMC). This leaves 2134 red giants
in the LMC and 957 in the SMC. Their distribution is shown in panels {\bf a}
and {\bf b} of Fig.~\ref{fig:apogee}, for the LMC and SMC, respectively,
and their metallicity values against distance from the LMC/SMC optical center
are plotted in panels {\bf c} and {\bf d}.
In the case of the SMC we adopt an elliptical system with the major
to minor axis ratio of $1.5$ and a position angle of $55.3^\circ$ east of north,
in order to account for the elongation of the SMC and to be consistent with
previous metallicity studies (e.g. \citealt{Piatti2007,Dobbie2014,Parisi2016}).
The line fit to the data gives a
metallicity gradient of $-0.026 \pm 0.002$ dex/deg in the LMC. In the majority
of the SMC, the gradient is consistent with $-0.033 \pm 0.005$ dex/deg, but there is a clear
change in the slope at about $1.5^\circ$ distance (in elliptical coordinates),
and the gradient can be described by $-0.118 \pm 0.027$ dex/deg in the central part of the
SMC.

Since the locations of the APOGEE fields are sparse across the galaxies,
and the scatter of [Fe/H] values is large, it is difficult to investigate the
metallicity gradient in more detail, i.e. verify whether it is uniform in all
radial directions. However, if the metallicity gradient was variable, it would
not significantly alter the results of this paper due to a very shallow
dependence of $(V-I)_0$ on [Fe/H] within the metallicity range of the MCs,
which is discussed below.

\vspace{0.1cm}

The relation between the intrinsic color of the RC and its metallicity
is supported by both theory and observations \citep{Girardi2016,Nataf2020}.
In an attempt to calibrate this dependence, we selected a sample of 17 star
clusters from the LMC and SMC, which are older than 3~Gyr and younger than
9.5~Gyr, as their RC color does not change with age over this age range (see
Fig.~1 of \citealt{Girardi2001}). The selection also required that the clusters
are located in areas with low reddening ($E(V-I\,)_{\rm SFD} < 0.12$~mag).
We use their [Fe/H] values available in the literature
\citep{DaCosta1998,Bica1998,Grocholski2006,Glatt2008,Parisi2009,Dias2016},
with the preference of spectroscopically determined metallicities
over the photometric ones, if both were available. The spectroscopically
determined metallicities are based on the calcium triplet technique and
have typical uncertainties of $\sim 0.05$~dex. The photometric estimations 
are based on comparing the observed and synthetic CMDs and have much larger
errors, usually $> 0.20$~dex.
The RC colors of the 17 clusters are measured using OGLE-IV data within the
cluster radius, with a typical error $< 0.02$~mag, and dereddened with the
SFD maps. Since MCs star clusters meeting these criteria are rather metal poor,
we supplement the above dataset with data for NGC~6791 and the solar
neighborhood \citep{Nataf2020} and for the Galactic Bulge \citep{Bensby2017}.

\vspace{0.1cm}
\begin{figure}[t]
\centerline{\includegraphics[width=0.48\textwidth]{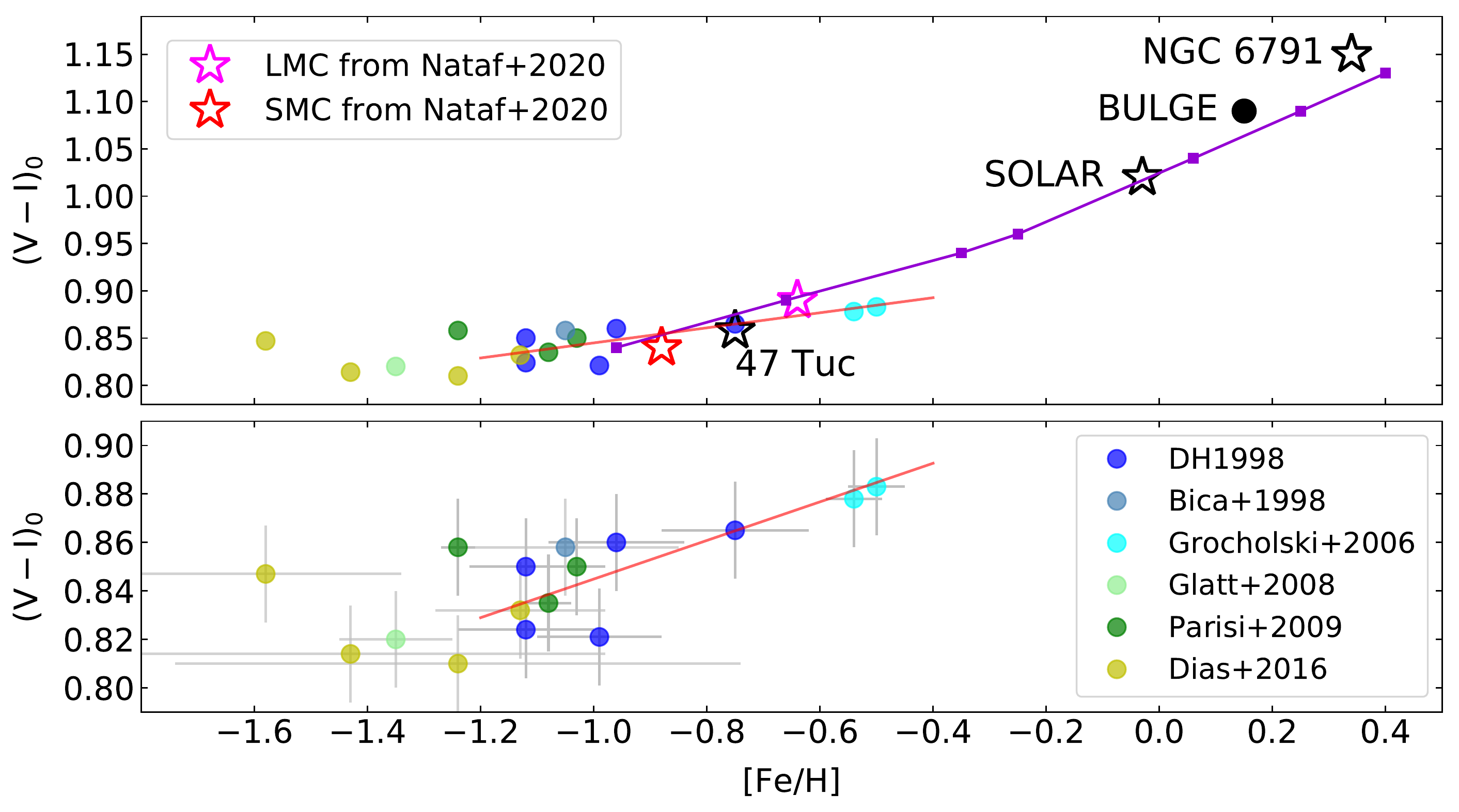}}
\caption{
The dependence of the intrinsic RC color on metallicity. The bottom
panel is a close-up of the top panel to show the measurement uncertainties
in the metallicity range considered in this work.
The cluster metallicity data is for clusters older than $3$~Gyr
and younger than $9.5$~Gyr,  on the \cite{Carretta1997} metallicity scale.
The data are from \cite{DaCosta1998}, \cite{Bica1998}, \cite{Grocholski2006},
\cite{Glatt2008}, \cite{Parisi2009} and \cite{Dias2016}.
The intrinsic RC color for all the above is based on OGLE-IV data.
Both [Fe/H] and $(V-I\,)_0$ for 47~Tuc, the solar neighborhood and NGC~6791
are from \cite{Nataf2020} and are marked with black stars (see their Fig.~1
for comparison). The data for the Galactic Bulge are from \cite{Bensby2017}.
For comparison, the results from \cite{Nataf2020} for the LMC and SMC
are marked with pink and red stars, respectively.
The purple line is the theoretical relation from \cite{Nataf2020}
shifted by $-0.03$~mag (see their Section 2.4).
The red line shows a line fit for $-1.2 < {\rm [Fe/H]} < -0.4$~dex, that
accounts for both [Fe/H] and $(V-I\,)_0$ uncertainties, with the slope of
$0.080 \pm 0.016$~mag/dex and the zeropoint of $0.925 \pm 0.016$~mag. 
}
\vspace{0.3cm}
\label{fig:clusters}
\end{figure}

\vspace{0.1cm}
\begin{figure*}[htb]
\centerline{\includegraphics[width=1.0\textwidth]{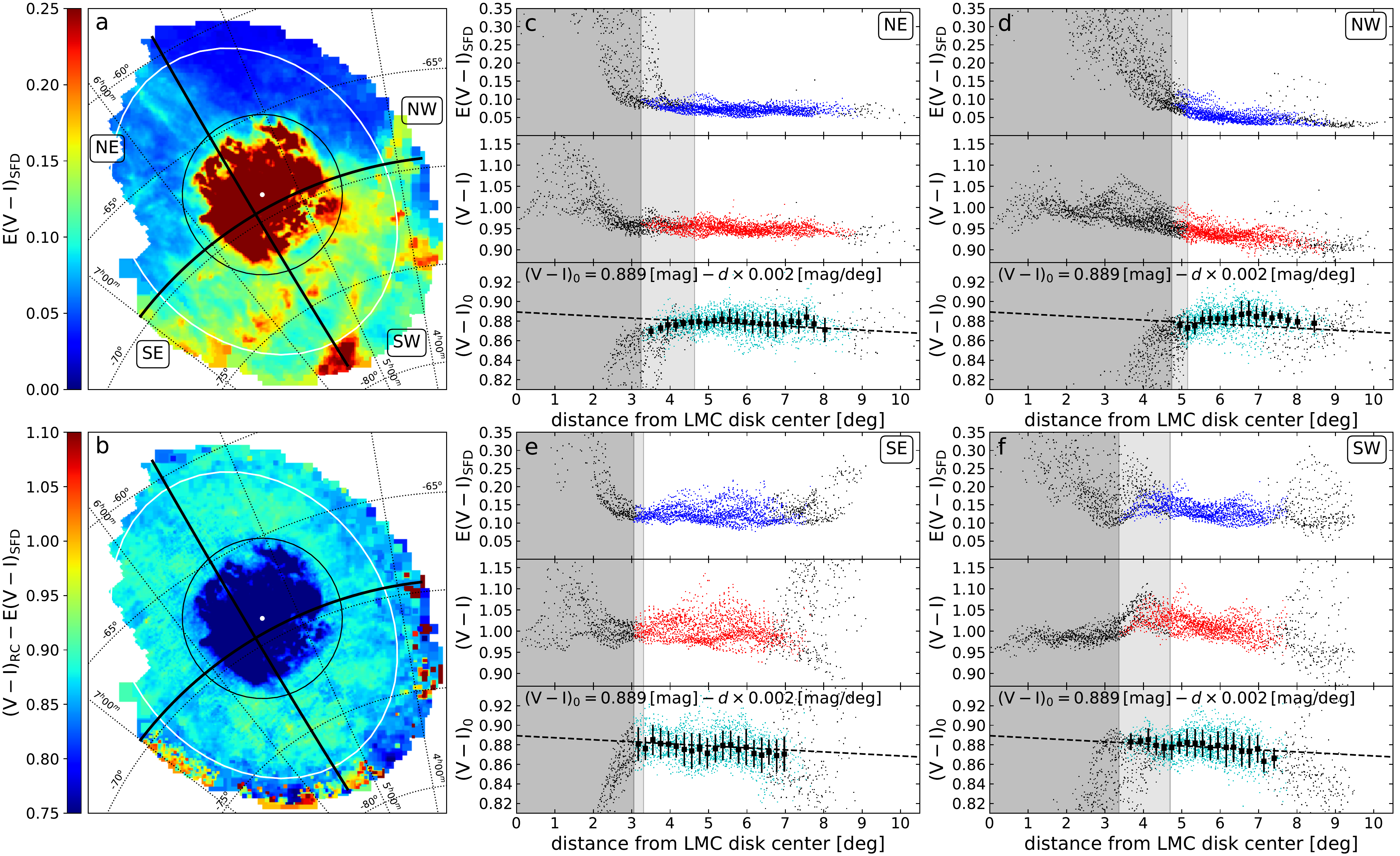}}
\caption{
The intrinsic color of the RC in the LMC.
The distribution of $E(V-I\,)_{\rm SFD}$ MW reddening in the LMC is shown in
panel~{\bf a}, while the difference between the measured $(V-I\,)$ color of the
RC and the MW reddening is shown in panel~{\bf b}. The circle marks the area
excluded from the analysis of $(V-I\,)_0$ and is centered at the LMC center of
rotation in radio and optical marked with a white dot \citep{Feast1961}.
The black solid lines of constant right ascension and declination cross at
the center of the LMC disk \citep{vanderMarel2001} and subdivide the LMC disk
into four regions: north-east, north-west, south-east, south-west. Panels
{\bf c-f} represent data for the four quadrants of the LMC, as marked in
panel {\bf a}. Each of the panels {\bf c-f} is composed of three plots:
the MW reddening $E(V-I\,)_{\rm SFD}$ (top, blue points), the measured RC color
$(V-I\,)$ (middle, red points) and their difference, i.e. the inferred
intrinsic RC color $(V-I\,)_0$ (bottom, cyan points), against the distance
from the LMC disk center. The dark-gray shaded area marks the central region
entirely excluded from the analysis, while the light-gray shaded area
marks the region partially excluded from the analysis (the black circle
crosses a range of distances because it is centered at different coordinates).
Colored points represent valid data, while black points show data
excluded from the analysis. Black squares with error bars (bottom plots in
panels {\bf c-f}) mark median $(V-I\,)_0$ values and their standard deviations
within bins of $0.2^{\circ}$. The black dashed line represents
a fit to cyan points that accounts for the fixed metallicity gradient taken
from panel {\bf c} of Fig.~\ref{fig:apogee}.
}
\vspace{0.3cm}
\label{fig:vi0_lmc}
\end{figure*}

Fig.~\ref{fig:clusters} shows the dependence of the intrinsic RC color
$(V-I\,)_0$ on metallicity [Fe/H] for the above sample.
The top panel includes the theoretical relation between $(V-I\,)_0$ and
[Fe/H] from \cite{Nataf2020}, which was limited to ${\rm [Fe/H]} \sim
-1.0$~dex at the low metallicity end (the purple line). The red line shows a line
fit to the datapoints, within the metallicity range considered in this work
($-1.2 < {\rm [Fe/H]} < -0.4$~dex). The estimated slope is $0.080 \pm
0.016$~mag/dex and the zeropoint is $0.925 \pm 0.016$~mag. The fit takes into
account both [Fe/H] and $(V-I\,)_0$ uncertainties, marked in the bottom panel
of Fig.~\ref{fig:clusters}.
The intrinsic color change in this metallicity range is nonnegligible but small
and the estimated slope may change with the availability of new cluster data.
The calculated slope of $0.08$~mag/dex, combined with the metallicity 
gradient\footnote{There is an offset of $\sim 0.06$~dex between the
metallicity scale of APOGEE and of Carretta et al. \citep{Nidever2020}, but it
is not relevant for this study as we are only using the slope of the relation,
not the zeropoint.}  in both galaxies (as estimated in Fig.~\ref{fig:apogee}),
predicts an intrinsic  color change with the distance from the galaxy center
of the order of $-0.002$~mag/deg in the LMC, $-0.003$~mag/deg in the outer SMC,
and $-0.009$~mag/deg in the inner SMC.

\vspace{0.1cm}
\begin{figure*}[t]
\centerline{\includegraphics[width=1.0\textwidth]{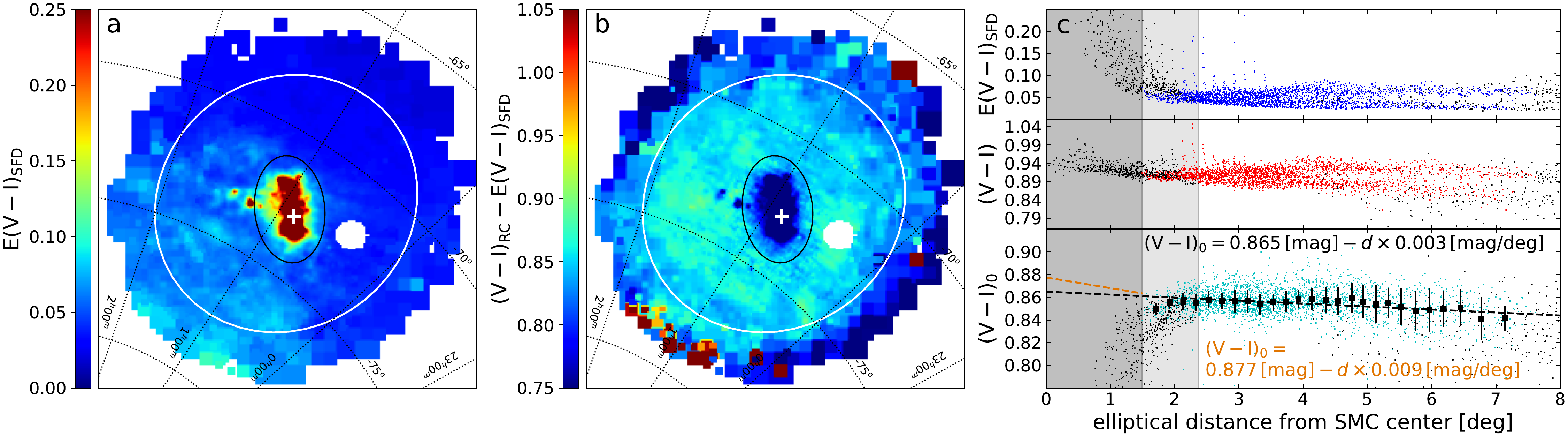}}
\caption{
The intrinsic color of the RC in the SMC.
The distribution of $E(V-I\,)_{\rm SFD}$ MW reddening in the SMC is shown in
panel~{\bf a}, while the difference between the measured median $(V-I\,)$
color of RC and the MW reddening is shown in panel~{\bf b}. The black ellipse
marks the area excluded from the analysis of $(V-I\,)_0$. Panel~{\bf c} is
composed of three plots: the MW reddening $E(V-I\,)_{\rm SFD}$ (top, blue
points), the measured RC color $(V-I\,)$ (middle, red points) and the inferred
intrinsic RC color $(V-I\,)_0$ (bottom, cyan points), against the elliptical
distance from the SMC disk center \citep{deVaucouleurs1972}, marked with a cross
in panel {\bf a}. The dark-gray shaded area marks the central region
entirely excluded from the analysis, while the light-gray shaded area
marks the region partially excluded from the analysis (the black ellipse
crosses a range of distances because it is centered at different coordinates).
Colored points represent valid data, while black points show data
excluded from the analysis. Black squares with error bars (bottom plot in
panel~{\bf c}) mark median $(V-I\,)_0$ values and their standard deviations
within bins of $0.2^{\circ}$. The black and orange dashed lines represent
a fit to cyan points that accounts for the fixed metallicity gradients from
panel {\bf d} of Fig.~\ref{fig:apogee}.
}
\vspace{0.3cm}
\label{fig:vi0_smc}
\end{figure*}

\subsection{$(V-I\,)_0$ of the Red Clump in the LMC}
\label{sec:vi0_lmc}

The distribution of $E(V-I\,)_{\rm SFD}$ reddening in the MW around the LMC is
presented again in panel~{\bf a} of Fig.~\ref{fig:vi0_lmc}, while panel~{\bf b}
shows the distribution of the unreddened RC color, i.e. the difference between
the measured median $(V-I\,)$ color of RC and $E(V-I\,)_{\rm SFD}$
reddening in the MW around the LMC.
The MW dust distribution in the LMC region is non uniform -- its southern part
is highly affected by dust, the north eastern part is moderately affected,
whereas the north western part is almost unaffected by Galactic extinction. We
therefore subdivide the LMC into four quadrants (Fig.~\ref{fig:vi0_lmc}, panels
{\bf a-b}), along constant right ascension and declination, that cross in the
center of the LMC disk ($05^{\rm{h}}29^{\rm{m}}00^{\rm{s}}$,
$-68^{\circ}30'00''$, from \citealt{vanderMarel2001}). Note that the center
of dust emission is offset from the center of the LMC disk by over one degree.
Each of the panels {\bf c-f} of Fig.~\ref{fig:vi0_lmc} contains three plots:
the MW reddening $E(V-I\,)_{\rm SFD}$ (top, blue points), the measured
median RC color $(V-I\,)$ (middle, red points) and their difference,
i.e. the inferred intrinsic RC color $(V-I\,)_0 = (V-I\,) - E(V-I\,)_{\rm SFD}$
(bottom, cyan points), against the distance from the center of the LMC disk.
They represent data within the four quadrants of the LMC: north-east
(panel~{\bf c}), north-west (panel~{\bf d}), south-east (panel~{\bf e}), and
south-west (panel~{\bf f}). Black points in all plots show spurious measurements
that either fall inside the black circle (where $E(V-I\,)_{\rm SFD}$ is wrong),
or outside the white ellipse (where RC color measurements are spurious) pictured
in panels {\bf a-b}. The dark-gray shaded area marks the central region
entirely excluded from the analysis, while the light-gray shaded area
marks the region partially excluded from the analysis (the black circle
crosses a range of distances due to the difference in coordinates of the LMC 
dust and stellar disks).

The line fit to $(V-I\,)_0$ {\it vs} distance from the LMC center (cyan points)
is shown with a dashed line in the bottom plots of panels {\bf c-f}. The fit is
based on joined data from the NE, SE and SW quadrants, and accounts for the
fixed metallicity gradient and the resulting RC color change of
$-0.002$~mag/deg. The different behavior of $(V-I\,)_0$ in the NW is unlikely
to be caused by the actual intrinsic color difference in this LMC quadrant, but
is very likely to be a systematic effect originating from the calibration of
SFD extinction maps. This effect becomes most prominent in LMC regions with
lowest MW extinction, probably because the contribution of the
LMC internal dust is higher than that of the MW foreground dust, for which
the relation between dust temperature and reddening was calibrated.
Another possibility is that $E(V-I\,)_{\rm SFD}$ is underestimated in regions
with very low dust density.
For these reasons, we did not use the NW quadrant, where the reddening is the
lowest, for fitting the intrinsic color.

The median value of $(V-I\,)_0$ in the investigated area, between the black
circle and the white ellipse, is $0.877$~mag. The scatter of individual lines of
sight around the fitted line is $0.014$~mag. 
The predicted intrinsic color in the LMC center from the color gradient is
$0.886$~mag, which is consistent with the results of \cite{Nataf2020}.

\begin{table*}
\caption{E(V-I\,) reddening map based on RC color from the OGLE-IV survey.\label{tab:map}}
\begin{tabular}{cccccccccc}
\hline
 RA   &    Dec      & $E(V-I\,)$ & $\sigma_1$ & $\sigma_2$ & $(V-I)_{\rm RC}$ & $(V-I)_0$ & $E(V-I\,)_{\rm peak}$ & $E(V-I\,)_{\rm SFD}$ & box \\
${\rm [hr]}$ & [deg] &  median   &  [-34\%]   &  [+34\%]   &     median       &           &      mode             &                      & [arcmin]  \\
\hline
4.866824 & -68.599136  &  0.128  &   0.046    &  0.057     &      1.010       &  0.882    &          0.127        &         0.357        &  3.4\\
5.434266 & -63.469700  &  0.066  &   0.061    &  0.044     &      0.943       &  0.877    &          0.074        &         0.061        &  6.9\\
0.379222 & -74.886131  &  0.039  &   0.030    &  0.028     &      0.896       &  0.857    &          0.042        &         0.041        & 13.8\\
   ...   &    ....     &   ....  &    ....    &   ....     &       ....       &   ....    &           ....        &          ....        & .... \\
\hline
\end{tabular}
\end{table*}

\subsection{$(V-I\,)_0$ of the Red Clump in the SMC}
\label{sec:vi0_smc}

The MW extinction around the SMC is lower and its distribution is more
uniform than around the LMC (see panel~{\bf a} of Fig.~\ref{fig:vi0_smc}),
with only a slight difference between the MW dust distribution in the northern
and southern parts of the SMC.
The distribution of the unreddened RC color, i.e. the difference between the
measured median $(V-I\,)$ color of RC and $E(V-I\,)_{\rm SFD}$ reddening
in the MW around the SMC is presented in panel~{\bf b} of Fig.~\ref{fig:vi0_smc}.
Similarly as in the case of the LMC, panel~{\bf c} shows the MW reddening
$E(V-I\,)$ (top, blue points), the measured RC color $(V-I\,)$ (middle, red
points) and the inferred intrinsic RC color $(V-I\,)_0$ (bottom, cyan points),
against the elliptical distance from the center of the SMC at
($00^{\rm{h}}52^{\rm{m}}12.5^{\rm{s}}$, $-72^{\circ}49'43''$) from
\cite{deVaucouleurs1972}, with the major to minor ellipse axis ratio of $1.5$
and a position angle of $55.3^\circ$ east of north
{\bf (e.g. \citealt{Piatti2007, Dobbie2014,Parisi2016}).}
Black points in all plots of panel~{\bf c} show spurious measurements that
either fall inside the black ellipse (where $E(V-I\,)_{\rm SFD}$ is wrong), or
outside the white ellipse (where RC color measurements are spurious) pictured
in panels {\bf a-b}. The dark-gray shaded area marks the central region
entirely excluded from the analysis, while the light-gray shaded area
marks the region partially excluded from the analysis (the black ellipse
crosses a range of distances from the SMC center).

The line fit to $(V-I\,)_0$ {\it vs} elliptical distance from the SMC center
(cyan points) is shown in the bottom plot of panel {\bf c} with a dashed line.
The fit accounts for the fixed metallicity gradient and the resulting RC color
change of $-0.003$~mag/deg for distances larger than $\sim 1.5^\circ$ (black
line) and $-0.009$~mag/deg for distances smaller than $\sim 1.5^\circ$ (orange
line). 

The median value of $(V-I\,)_0$ in the investigated area, between the black
and white ellipses is $0.853$~mag. 
The scatter of individual measurements around the fitted line is $0.018$~mag.
The predicted intrinsic color in the SMC center is $0.877$~mag, which is much higher
than $0.84$~mag estimated by \cite{Nataf2020} and is a result of the steepening
of the metallicity gradient in the inner SMC. If we assumed that the gradient 
does not change and follows the black dashed line (panel {\bf c} of
Fig.~\ref{fig:vi0_smc}), then the predicted intrinsic color in the SMC center
would be $0.865$~mag.
Another reason for the high central value of $(V-I\,)_0$ may be due to the
underestimation of $E(V-I\,)_{\rm SFD}$ in areas with very low reddening.
Similarly as in the NW quadrant of the LMC, where increasing
$E(V-I\,)_{\rm SFD}$ by a few hundredths would shift the reddening estimates
to fit the $(V-I\,)_0$ gradient estimated from other quadrants,
in the SMC (where the extinction is generally much lower than in the LMC), it
would cause the lowering of $(V-I\,)_0$.

\section{Reddening Maps}
\label{sec:maps}

\vspace{0.1cm}
\begin{figure*}[htb]
\centerline{\includegraphics[width=0.8\textwidth]{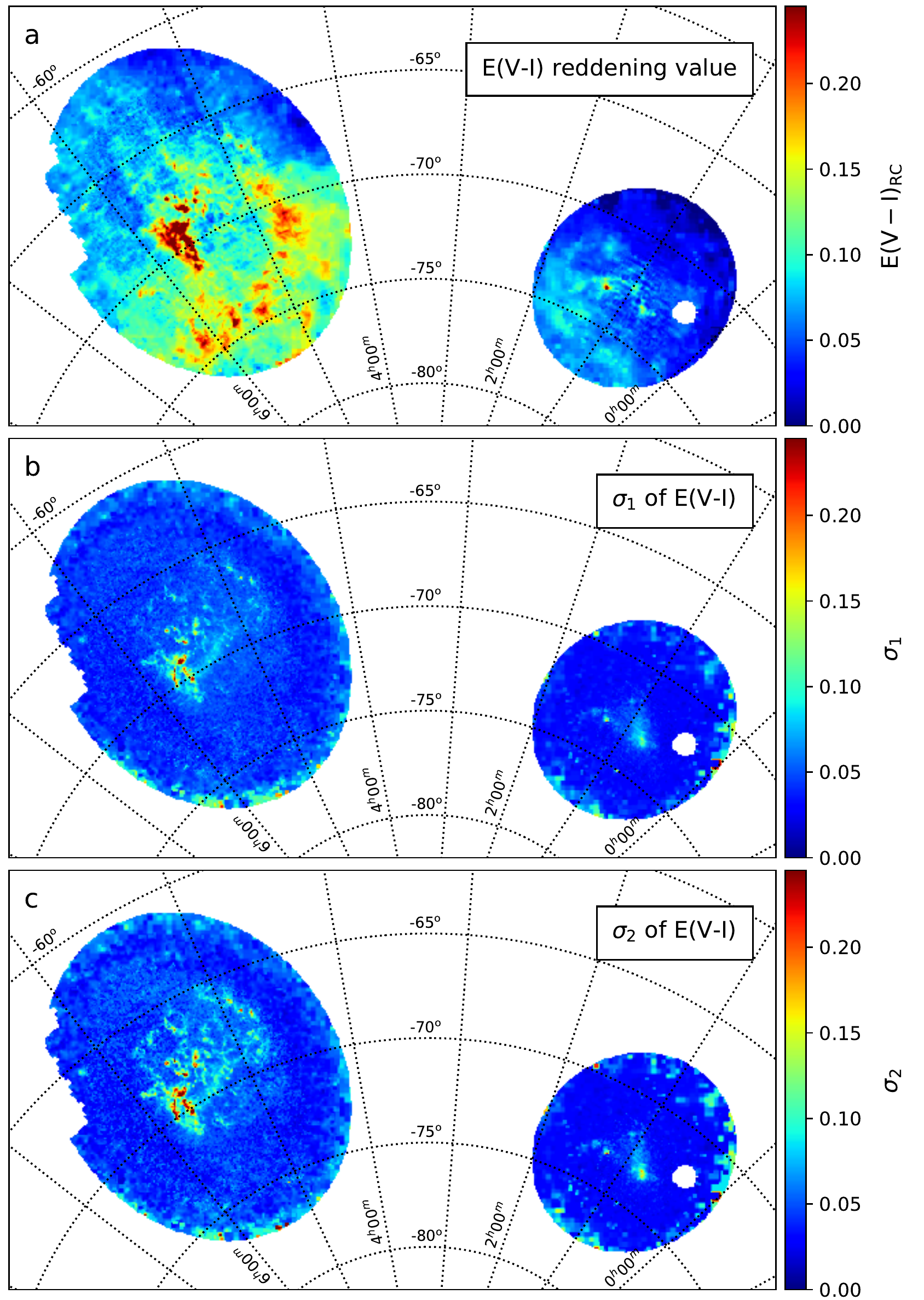}}
\caption{
The reddening map of the Magellanic Clouds. Panel~{\bf a} shows the distribution
of $E(V-I\,)$ reddening in the MCs. Panels {\bf b} and {\bf c} show the
distribution of $\sigma_1$ and $\sigma_2$, respectively, which are an indication
of the amount of dust in the near parts ($\sigma_1$) and in the central and far
parts ($\sigma_2$) of the galaxies, and can serve as measures of reddening
uncertainties toward a single star. The empty region within the SMC is the
location of the globular cluster 47~Tuc.
}
\vspace{0.3cm}
\label{fig:final_maps}
\end{figure*}

The final $E(V-I\,)$  reddening map of the MCs was obtained by subtracting
the intrinsic RC color (Sections~\ref{sec:vi0_lmc} and~\ref{sec:vi0_smc}) from
the measured one. The resolution of the map is $1.7' \times 1.7'$ in the
central parts of the MCs and decreases down to approximately $27' \times 27'$
in the outskirts.

Fig.~\ref{fig:final_maps} shows
the distribution of reddening (panel~{\bf a}) and its standard deviation:
$\sigma_1$ (panel~{\bf b}) and $\sigma_2$ (panel~{\bf c}). 
Both $\sigma_1$ and $\sigma_2$ are measures of internal reddening
in the MCs and inform of the distribution of dust within the galaxies
along the line of sight -- high $\sigma_1$ indicates increased amounts of dust
in the near parts, while high $\sigma_2$ in the central and far parts of the
galaxy. 
It is well represented in panel~{\bf c}, where $\sigma_2$ reflects the 
distribution of dust within the LMC inner disk (see Section~\ref{sec:rccolor}
for details).

The mean reddening is $E(V-I\,) = 0.100 \pm 0.043$~mag in the LMC and $E(V-I\,)
= 0.047 \pm 0.025$~mag in the SMC. This is lower than all previous measurements,
but is expected -- this is the first reddening map of the MCs that encompasses
all areas where the RC is visible, including the outskirts of the galaxies,
where the reddening is very low. 

The reddening data are available on-line from:

\vspace{0.1cm}
\hspace{0.5cm} {\it http://ogle.astrouw.edu.pl/cgi-ogle/get\_ms\_ext.py}
\vspace{0.1cm}

\noindent both for download (in TEXT and FITS formats, for the users convenience) and in
the form of an interactive interface. The subset of the data is presented in
Table~\ref{tab:map}.

\subsection{Comparison with Previous Large Scale Reddening Maps Based on RC Stars.}

Here we compare our reddening maps with other large scale reddening maps of
the Magellanic Clouds based on RC stars. 
Instead of comparing reddening values themselves, which highly depend on the
adopted intrinsic color of the RC, we rather compare maps of measured
RC color and the adopted intrinsic RC color separately.

\subsubsection{Haschke et al. (2011)}

\vspace{0.1cm}
\begin{figure}[th]
\centerline{\includegraphics[width=0.48\textwidth]{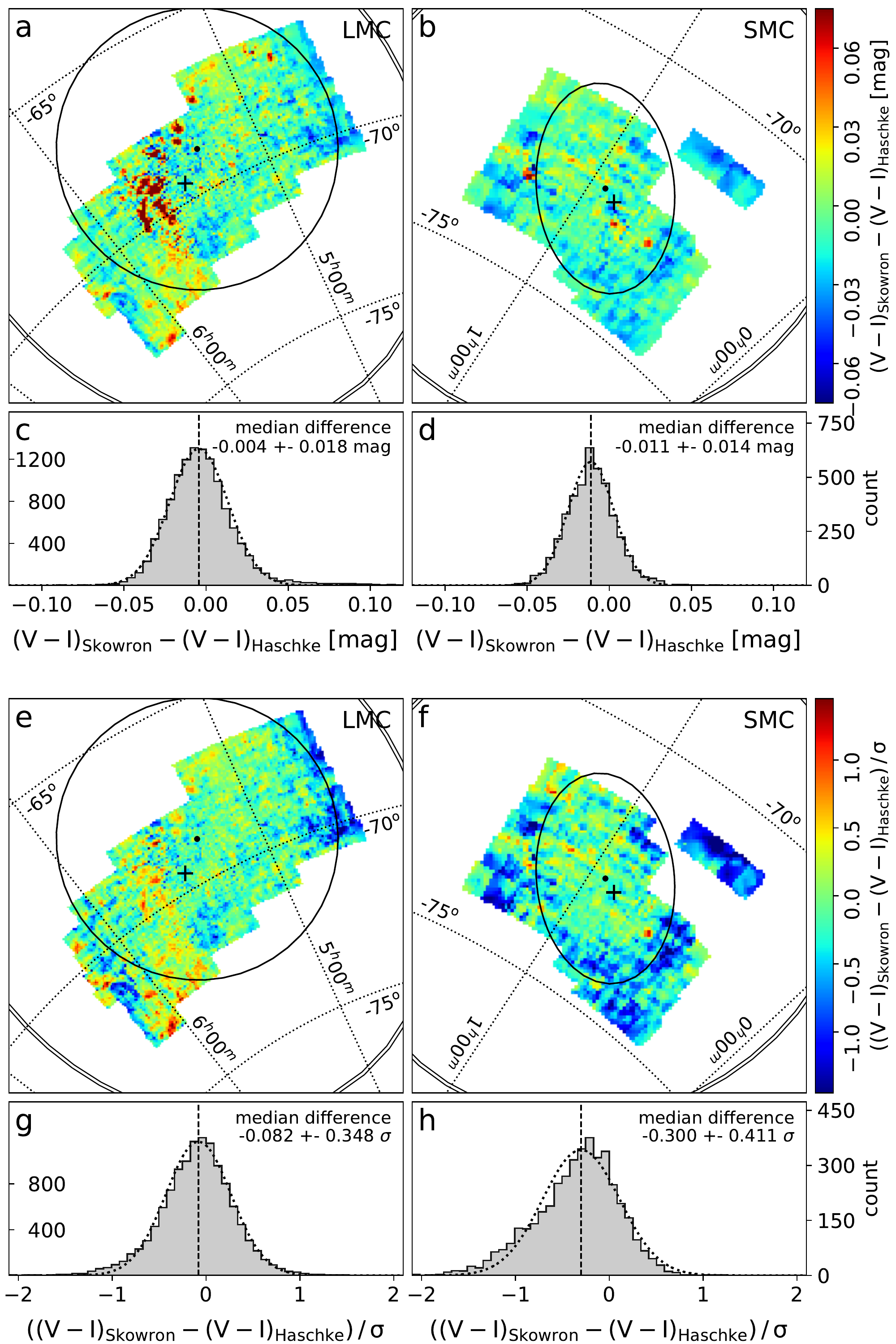}}
\caption{
The comparison of color measurements from this work and the work of
\cite{Haschke2011} as a two-dimensional map of RC color differences
in the LMC (panel~{\bf a}) and SMC (panel~{\bf c}), and as a histogram of
color differences (panels {\bf b} and {\bf d}).
Panels {\bf e} and {\bf f} show the two-dimensional map of
RC color differences divided by the $\sigma$ of the color distribution for
the LMC and SMC, respectively, while panels {\bf g} and {\bf h} show the
corresponding histograms.
}
\vspace{0.3cm}
\label{fig:compare-haschke}
\end{figure}

Large scale reddening maps of the LMC and SMC published by \cite{Haschke2011}
were based on RC stars from the OGLE-III data \citep{Udalski2008a,Udalski2008b}.
Authors found a mean reddening $E(V-I\,) = 0.09 \pm 0.07$~mag in the LMC and
$E(V-I\,) = 0.04 \pm 0.06$~mag in the SMC. When taking into account only the
area investigated by \cite{Haschke2011}, we find mean reddening values of
$E(V-I\,) = 0.122 \pm 0.048$~mag in the LMC and $E(V-I\,) = 0.056 \pm 0.020$~mag
in the SMC. The difference is 0.032 and 0.016~mag for LMC and SMC, respectively, 
and is mostly a result of the adopted zero color of the RC. \cite{Haschke2011}
used theoretical values that depend on metallicity, adopting
$(V-I\,)_0 = 0.92$~mag and $(V-I\,)_0 = 0.89$~mag for the LMC and SMC,
respectively \citep{Girardi2001}, while we use empirical values, which together
with accounting for a metallicity gradient give mean intrinsic colors within
the area investigated by \cite{Haschke2011} of
$(V-I\,)_0 = 0.884$ (LMC) and $(V-I\,)_0 = 0.862$~mag (SMC).

Fig.~\ref{fig:compare-haschke} presents a comparison between color measurements
from this work and the work of \cite{Haschke2011}. Panels {\bf a} and {\bf b} 
show a two-dimensional map of RC color differences that are color coded. The
histogram of differences between $(V-I\,)$ color values from this work and from
\cite{Haschke2011} is shown in panel~{\bf c} for the LMC and in panel~{\bf d}
for the SMC. In the case of the LMC the mean difference is 
$\langle(V-I\,)_{Skowron} - (V-I\,)_{Haschke}\rangle = -0.004 \pm 0.018$~mag.
There are some expected larger discrepancies in the central region of the
LMC, where the differential reddening  is high, but the overall agreement is
reasonable and falls within one $\sigma$ value. This is further pictured in
panel {\bf e}, where a comparison of color differences divided by the $\sigma$
of the color is shown.

In the case of the SMC, there is a mean difference of
$\langle(V-I\,)_{Skowron} - (V-I\,)_{Haschke}\rangle = -0.011 \pm 0.014$~mag,
but there is also a systematic difference between the eastern and western parts,
due to small offsets between the OGLE-III and OGLE-IV calibrations (panel {\bf b}).
As seen in panel {\bf f}, there is an agreement within one $\sigma$ value for
the majority of the area, although larger differences are observed at the
edges, and these result from the differences in an algorithm calculating $(V-I\,)$.

\subsubsection{Choi et al. (2018)}

\begin{figure}[]
\centerline{\includegraphics[width=0.45\textwidth]{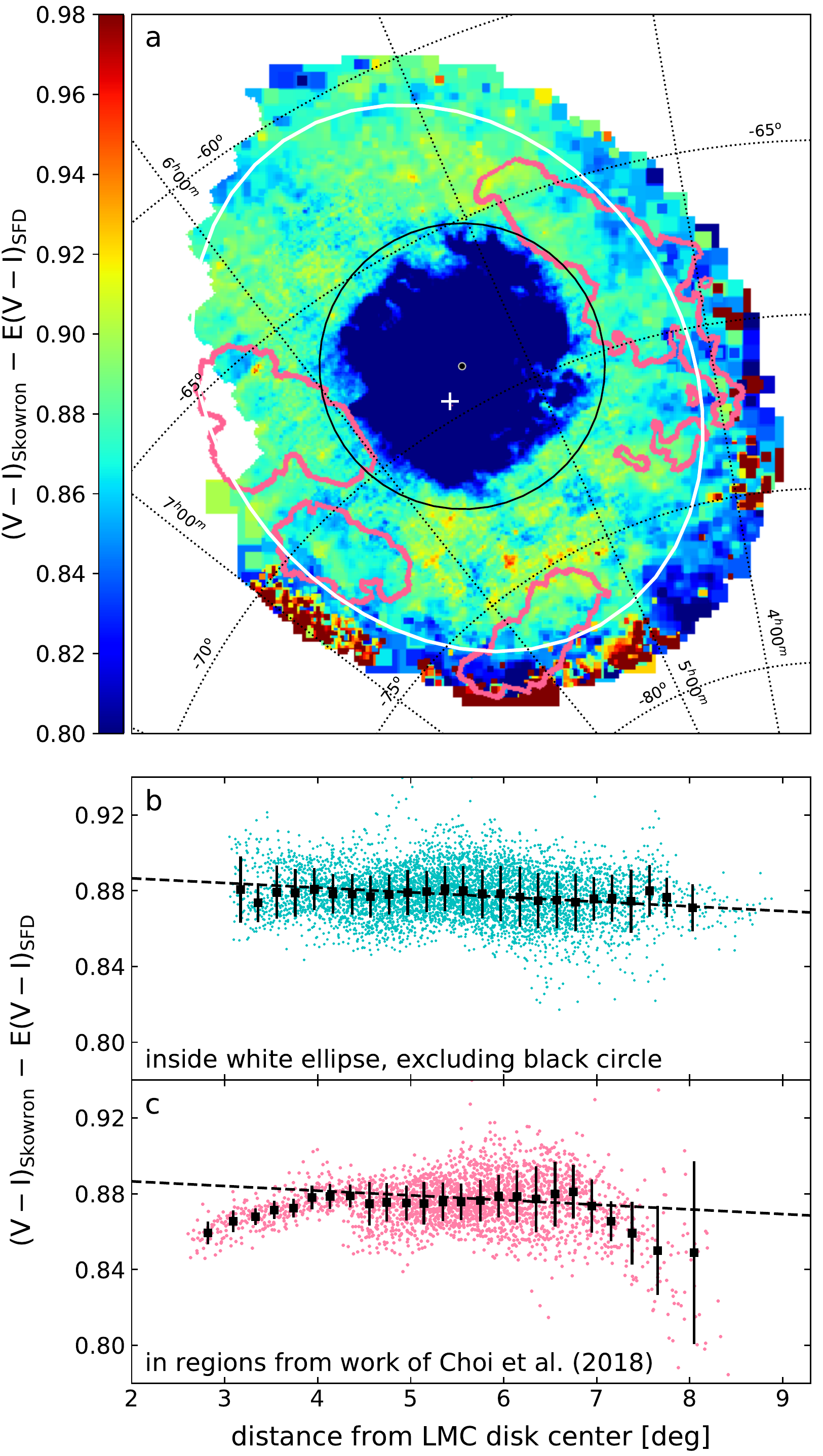}}
\caption{
The $(V-I\,)_0$ distribution of LMC RC stars from this work is shown in panel
{\bf a}. The black circle marks the central region of the LMC, where the SFD
map is not reliable, while the white ellipse cuts off the rejected outer region
of the galaxy, unsuitable for RC color measurements. Pink contours mark regions
used by \cite{Choi2018} to determine the intrinsic RC color, and are identical
as in their Fig.~4.
The plots in panels {\bf b} and {\bf c} show the change of $(V-I\,)_0$ measured
in this work, with distance from the LMC disk center (white cross). Panel~{\bf b}
(cyan points) shows $(V-I\,)_0$ in the "safe" region, between the black circle
and the white ellipse, while panel~{\bf c} (pink points) shows $(V-I\,)_0$
within pink contours.
The dashed line in panels {\bf b} and {\bf c} shows our derived $(V-I\,)_0$
gradient.
}
\vspace{0.3cm}
\label{fig:vi0_lmc_choi}
\end{figure}

More recently, \cite{Choi2018} released $E(g-i\,)$ reddening map of a large part
of the LMC based on RC stars observed by the Survey of the MAgellanic Stellar
History (SMASH, \citealt{Nidever2017}). They postulated that there is a radial
dependence of the intrinsic color of the RC, such that it is practically
constant in the range between $4^\circ$ and $7^\circ$ from the galaxy center,
with $(g-i\,)_0=0.822$, but follows a slope of $0.024$~dex/deg between
$2.7^\circ$ and $4^\circ$ and a slope of $-0.033$~dex/deg between
$7^\circ$ and $8.5^\circ$ from the center of the LMC (see their Fig.~7).
This finding is not supported by our data.
Panel~{\bf a} of Fig.~\ref{fig:vi0_lmc_choi} shows the distribution of $(V-I\,)_0$
from this work. The black circle marks the central region of the LMC, where
the SFD map is not reliable, while the white ellipse cuts off the rejected outer
region of the galaxy, unsuitable for color measurements.
In Section~\ref{sec:vi0} we demonstrated that only the region between the black
circle and white ellipse can be reliably used for determining the intrinsic color
of the RC. In panel~{\bf a} of Fig.~\ref{fig:vi0_lmc_choi} we use pink contours
to overplot regions used by \cite{Choi2018} to determine the intrinsic RC color,
which are identical as in their Fig.~4. It can immediately be seen that the
pink contours in the southern and western part of the LMC partially cover the 
outskirts of the LMC, where the $(V-I\,)$ determination is not reliable. 
Similarly, eastern and western contours close to the galaxy center enter the
region within the black circle, in which SFD reddening values, and so the 
$(V-I\,)_0$ determination, are not realistic. 

\vspace{0.1cm}

The plots in panels {\bf b} and {\bf c} of Fig.~\ref{fig:vi0_lmc_choi} show
the change of $(V-I\,)_0$ measured in this work, with distance from the LMC
disk center. Panel~{\bf b} (cyan points) presents intrinsic color values
in the "safe" region, between the black circle and the white ellipse, while
panel~{\bf c} (pink points) shows intrinsic color values within pink contours
used by \cite{Choi2018}. It is apparent that $(V-I\,)_0$ gradient in the range
$2.7^\circ - 4^\circ$ and  $7^\circ - 8.5^\circ$, visible in panel~{\bf c},
originates solely from the  choice of regions to determine $(V-I\,)_0$. However,
when we use data from the reliable area only, we only see a slight gradient
of $(V-I\,)_0$ in the entire range $3.5^\circ - 9^\circ$ from the LMC center
(panel~{\bf b}), originating from the metallicity gradient in the LMC.

When working with the reddening map of \cite{Choi2018}, which is provided in
the FITS format, we found that their map departures from the WCS convention.
The FITS-WCS standard (\citealt{Greisen2002}, Section 2.1.4)
states that the ``integer pixel numbers refer to the center of the pixel in
each axis, so that, for example, the first pixel runs from pixel number 0.5 to
pixel number 1.5 on every axis''. From our tests it appears that reddening
values of \cite{Choi2018} are for the area of
the sky between the coordinates calculated for the integer pixel numbers, e.g.
the first pixel represents area between WCS coordinates calculated for
positions 1 and 2. This half-pixel offset in both axes translates to the
typical offset of 7 arc minutes in their published map, and is clearly visible
when comparing with our maps, especially in the center, where the resolution
is $1.7' \times 1.7'$. We account for this offset in further
analysis\footnote{\cite{Choi2018} intend to correct their map
to conform with the FITS-WCS standard (private communication).}. 

In order to compare RC color values from this work and from \cite{Choi2018}
we use the provided $(g-i\,)$ color maps and transform them to $(V-I\,)$ with:
\begin{equation}
(V-I\,)_{\rm Choi} =  0.765\, (g-i\,) + 0.242
\end{equation}
The constant $0.242$ was chosen arbitrarily so that the mean color difference
is zero, since the exact value of the constant term is not known without a
detailed comparison of individual stars in both samples.

\vspace{0.1cm}
\begin{figure}[t]
\centerline{\includegraphics[width=0.44\textwidth]{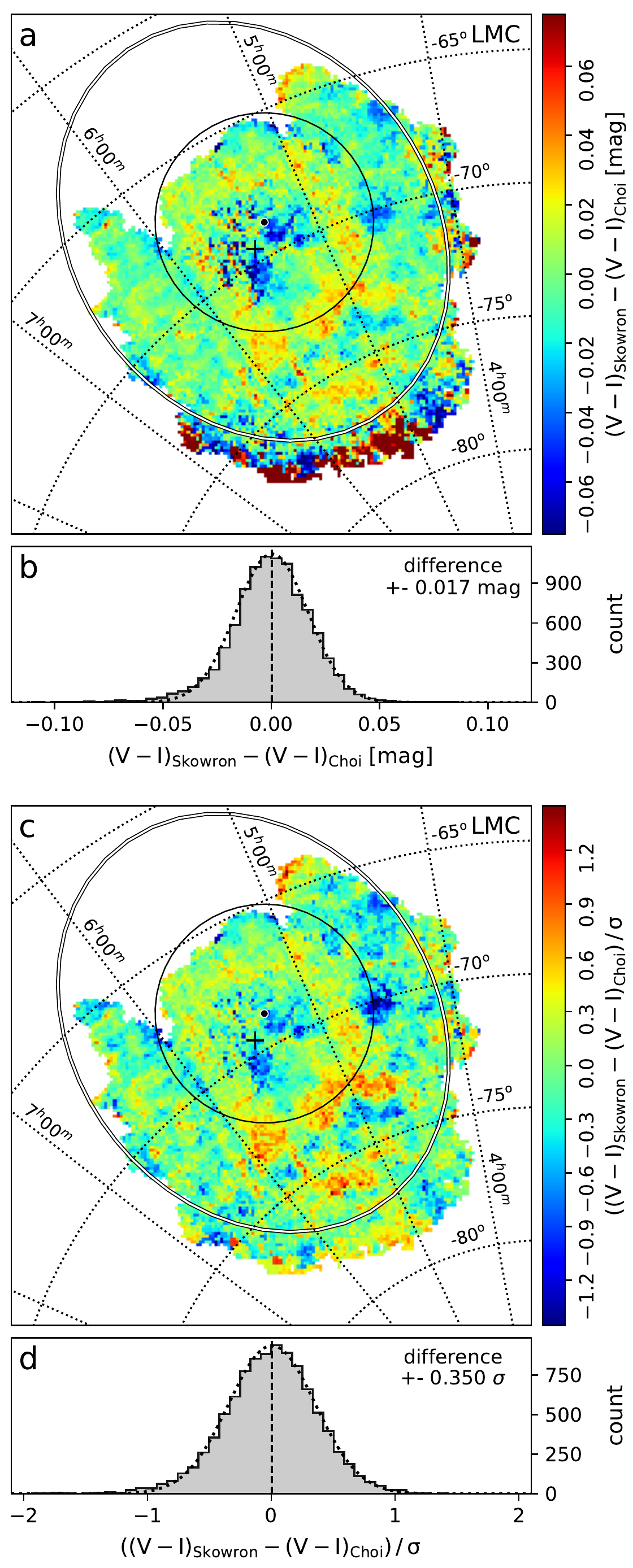}}
\caption{
The comparison of color measurements from this work and the work of
\cite{Choi2018} as a two-dimensional map of RC color differences
(panel~{\bf a}) and as a histogram of color differences (panels {\bf b})
in the LMC.
Panel {\bf c} shows the two-dimensional map of RC color differences
divided by the $\sigma$ of the color distribution, while panel {\bf c} shows
the corresponding histogram.
}
\vspace{0.3cm}
\label{fig:compare-choi}
\end{figure}

In Fig.~\ref{fig:compare-choi} we plot a color-coded 2D map (panel~{\bf a})
and a histogram (panel~{\bf b}) of differences between $(V-I\,)$ measured in
this work and obtained by \cite{Choi2018}. Panels {\bf c} and {\bf d} show a
comparison of color differences divided by the $\sigma$ of the color
distribution.
Even though there is a moderately good overall agreement between the two maps,
there are multiple regions with significant differences, which originate from
\cite{Choi2018} adopting the median value of the color distribution of all
stars in the selection box, as the RC color. This is especially seen in areas
with the higher scatter in color, i.e. in the center and in the outskirts. 
We investigated the nature of this discrepancy
by looking by eye at the CMDs of regions with the largest color differences.
We found that in these cases, the measurements of \cite{Choi2018} fall either
on the blue or on the red side of the actual RC color peak. This is caused
by the manual selection of the box containing RC stars. In the case when 
the RC is in the region of low extinction, their selection box omits some
of the RC stars on the red side, when trying to minimize the RGB contribution.
This is well pictured in Fig.~3 of \cite{Choi2018} -- the red part of the top
histogram in panel~{\bf a} is quite sharply cut off. The median of such
distribution will underestimate the real color of the RC, falling on the
blue side of the maximum of the RC color histogram.
In the other case, when the RC is in the region of high extinction, their
selection box contains mostly all RC stars on the red side, but there is
a sharp cut-off on the blue side in order not to include other
features that might contain stars from different populations.
(see the red part of the top histogram in panel~{\bf b} in Fig.~3 of
\cite{Choi2018}).
The median of such distribution will overestimate the real RC color,
falling on the red side of the maximum of the RC color histogram.

\subsubsection{G\'orski et al. (2020)}

Most recently, \cite{Gorski2020} published an update of the reddening maps of
\cite{Haschke2011} based on \mbox{OGLE-III} RC stars, with a new calibration
of the intrinsic RC color based on independently measured reddening of late-type
eclipsing binary systems, blue supergiants and reddening derived from Str\"omgren
photometry of B-type stars. They calculated the intrinsic RC color separately for
each tracer and then averaged the results to find  $(V-I\,)_0 = 0.838 \pm 0.034$
in the LMC and $(V-I\,)_0 = 0.814 \pm 0.034$ in the SMC.
This is disturbingly lower that the values obtained in this work.
If we average our intrinsic color estimates within the region investigated
by \cite{Gorski2020} we obtain mean $(V-I\,)_0$ of $0.883$~mag for the LMC and
$0.862$~mag for the SMC. The difference is  $0.045$ and $0.048$~mag for the LMC
and SMC, respectively. 
While it is true that our estimated central value of $(V-I\,)_0$ is based on
the RC color measurements in the outer disk, beyond $3^\circ$ from the LMC
center, while $(V-I\,)_0$ of \cite{Gorski2020} is based on the central part of
the LMC, the difference is nevertheless larger than expected. Since we
have accounted for the metallicity, and hence color gradient in the LMC, 
it is unlikely that our estimated intrinsic color in the inner parts of the
LMC differs by as much as $0.045$~mag.

\vspace{0.1cm}
\begin{figure}[t]
\centerline{\includegraphics[width=0.48\textwidth]{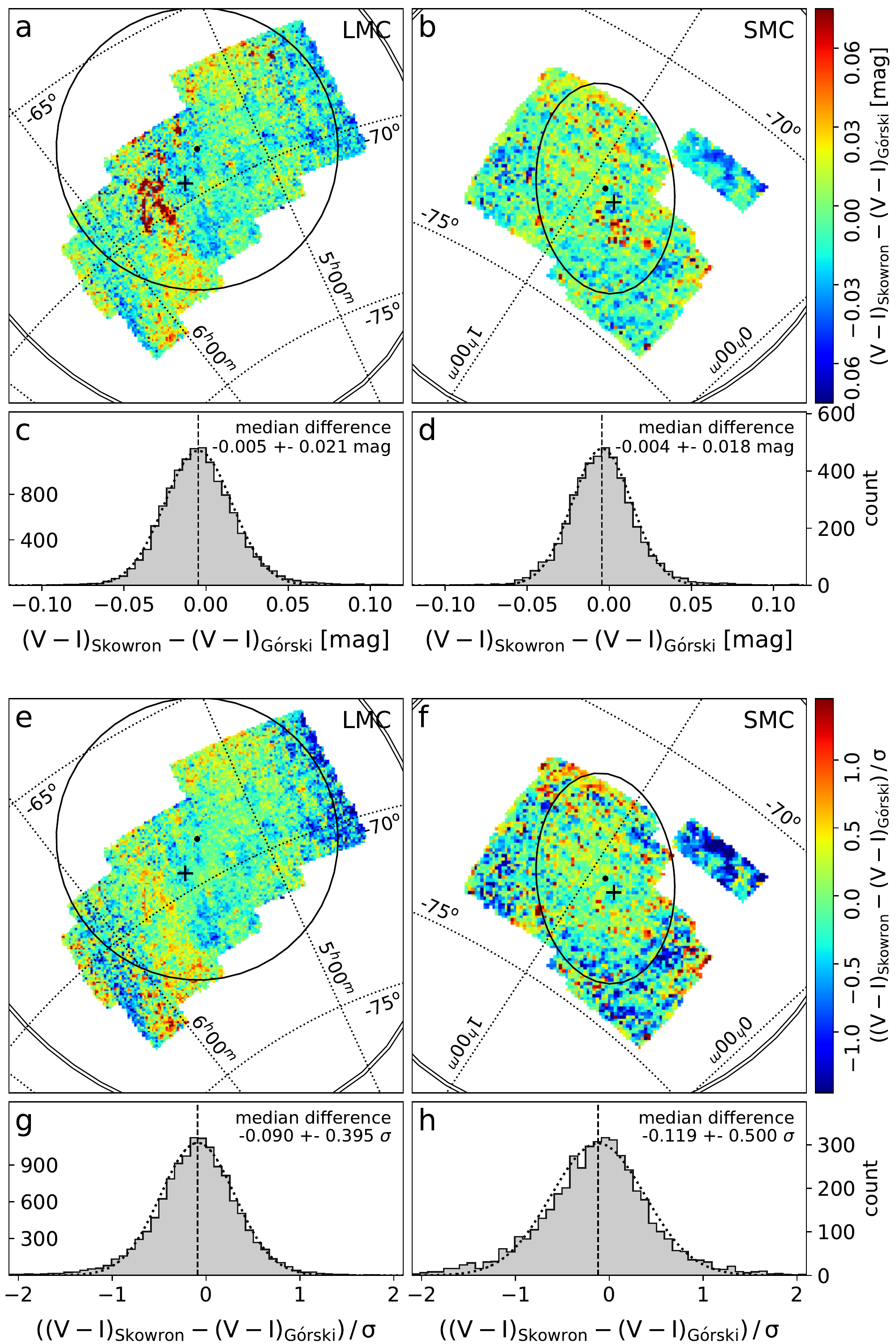}}
\caption{
The comparison of color measurements from this work and the work of
\cite{Gorski2020} as a two-dimensional map of RC color differences
in the LMC (panel~{\bf a}) and SMC (panel~{\bf c}), and as a histogram
of color differences (panels {\bf b} and {\bf d}).
Panels {\bf e} and {\bf f} show the two-dimensional map of
RC color differences divided by the $\sigma$ of the color for the
LMC and SMC, respectively, while panels {\bf g} and {\bf h} show the
corresponding histograms.
}
\vspace{0.3cm}
\label{fig:compare-gorski}
\end{figure}

This is further supported by a comparison of RC color measurements 
in this work and in \cite{Gorski2020}. Fig.~\ref{fig:compare-gorski} presents
a comparison of measured $(V-I\,)$ in the MCs, where panels {\bf a} and {\bf c}
show a two-dimensional map of differences that are color coded.
The histogram of differences between $(V-I\,)$ color values from this work
and from \cite{Gorski2020} is shown in panel~{\bf b} for the LMC and in 
panel~{\bf d} for the SMC. In the case of the LMC the mean difference is 
only $-0.005 \pm 0.021$ and in the SMC it is $-0.004 \pm 0.018$,
which shows that color estimates are consistent. 
Significant discrepancies in the galaxy centers are largely due to large
differential reddening and disappear in panels {\bf e} and {\bf f}, where the
RC color differences are  divided by the $\sigma$ of the color distribution. On the other
hand, panels {\bf e-f} highlight discrepancies in areas where the different
approach to estimating the RC color plays a role.

Tables~5 and~6 in \cite{Gorski2020} list separate $(V-I\,)_0$ values for each
of the tracers used to estimate the final intrinsic color. The method using
Na~I line in eclipsing binaries (from \citealt{Graczyk2014,Graczyk2018}) has
the smallest standard deviation of all four methods and gives consistent results
in terms of the intrinsic color difference between LMC and SMC of $0.023$~mag,
which is a reasonable number and is confirmed with our data.
Remaining methods tend to show smaller values of $(V-I\,)_0$, which can be
caused either by the additional reddening or a different reddening law in the
case of blue supergiants, or by systematic errors affecting reddening
determinations from effective temperature for eclipsing binaries or
Str\"omgren photometry (G\'orski M., private communication). 
It is also worth mentioning that the standard deviation of $(V-I\,)_0$ obtained
from the effective temperature of eclipsing binaries and atmospheric models of
blue supergiants is higher compared to the standard deviation of $(V-I\,)_0$
obtained from Na~I line. Additionally, the intrinsic color difference between
the LMC and SMC obtained from effective temperature of eclipsing binaries is
zero, for blue supergiants it is only $0.01$ and is as much as $0.068$~mag for
the Str\"omgren photometry, which is inconsistent with values obtained from Na~I
line analysis and our results. Thus, none of these values seem realistic, and
combined with high standard deviation of these methods, seem untrustworthy.
Given that the method using Na~I line in eclipsing binaries gives most reliable
results, one can assume, that the intrinsic color in the MCs obtained by
\cite{Gorski2020} should be rather $0.854$ and $0.831$, than $0.838$ and $0.814$,
in the LMC and SMC, respectively. This is closer to our $(V-I\,)_0$ averaged
within the region investigated by \cite{Gorski2020}, with the difference of
$0.029$ in the LMC and $0.031$ in the SMC.

The remaining question is whether this difference between the intrinsic RC
color in the outer (this work) and inner (work of \citealt{Gorski2020})
MCs, is of physical nature or is some sort of a statistical/calibration effect.
The argument in favor of it being real is that we observe a dip in $E(V-I\,)$
in the most central part of the LMC -- see panel~{\bf a} in 
Fig.~\ref{fig:final_maps}, where the color of the densest regions of the LMC
bar is bluer than the surrounding area, which may indicate a change in the RC
intrinsic color.

\vspace{0.1cm}
\begin{figure}[t]
\centerline{\includegraphics[width=0.48\textwidth]{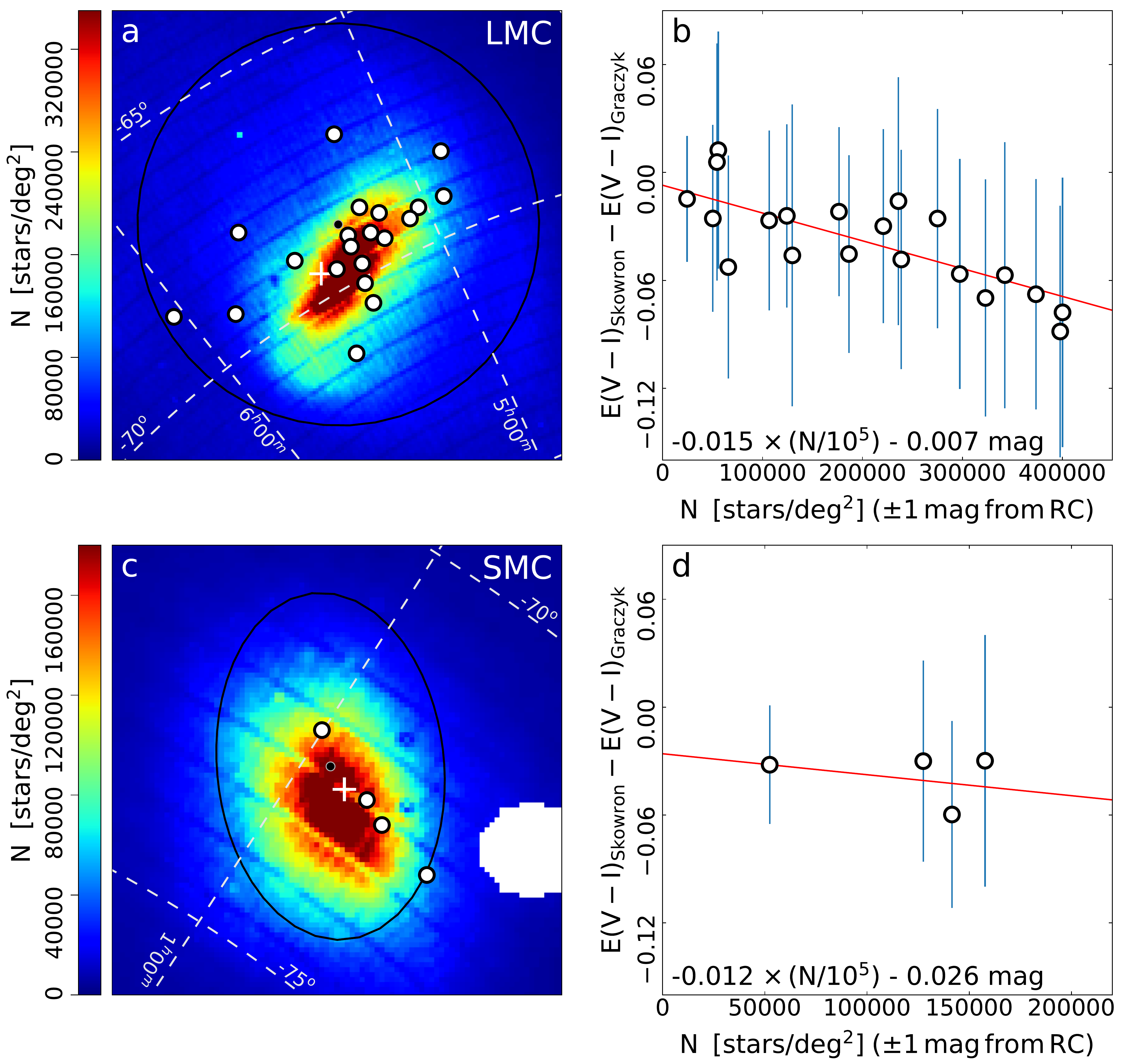}}
\caption{
The distribution of eclipsing binaries from \cite{Graczyk2014,Graczyk2018}
on top of the density map of stars within $\pm 1$~mag from the RC brightness
in the LMC (panel~{\bf a}) and SMC (panel~{\bf c}).
The $E(V-I\,)$ reddening difference between this work and the
work of \cite{Graczyk2014,Graczyk2018}, against the surface density
is presented in panels {\bf b} and {\bf d} for the LMC and SMC,
respectively.
}
\vspace{0.3cm}
\label{fig:compare-graczyk}
\end{figure}

\vspace{0.1cm}
\begin{figure*}[htb]
\centerline{\includegraphics[width=0.92\textwidth]{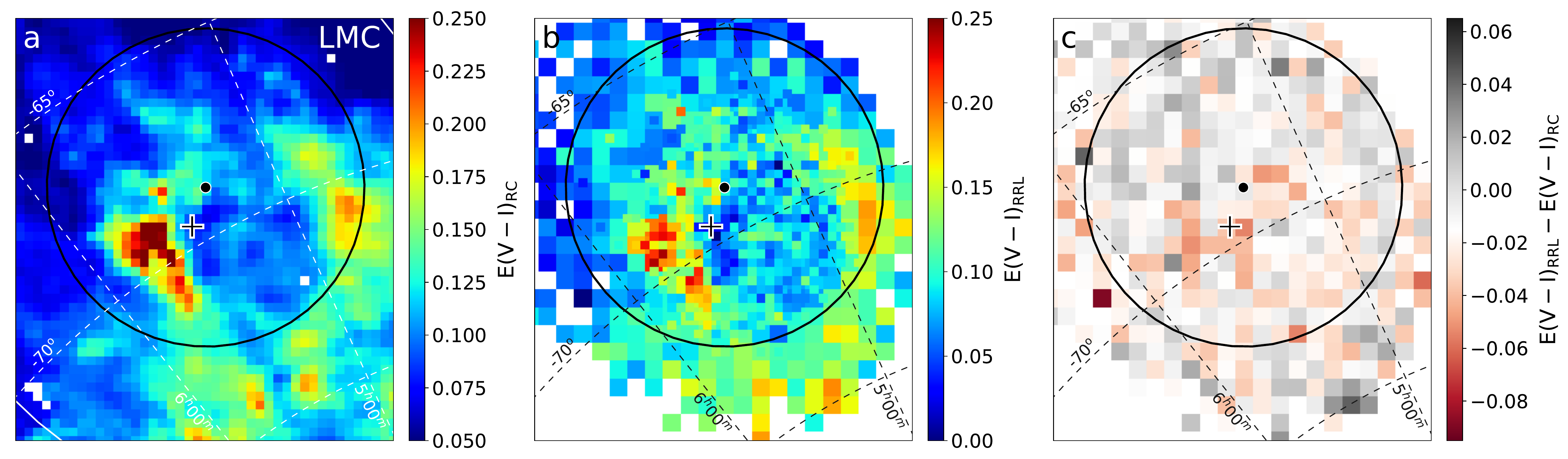}}
\caption{
The comparison of $E(V-I\,)$ reddening from RC and RR Lyrae stars in the LMC.
The reddening map of the LMC center based on RC stars is shown in panel~{\bf a}.
Panel~{\bf b} presents $E(V-I\,)$ in the same area based on RR Lyrae stars,
while panel~{\bf c} shows the difference of panels~{\bf a} and~{\bf b}.
}
\vspace{0.3cm}
\label{fig:compare-rrlyr}
\end{figure*}

After verifying that our RC color measurements are consistent with those of
\cite{Gorski2020}, we also verified our RC color measurements at the locations
of eclipsing binaries from \cite{Graczyk2014,Graczyk2018} and found very similar
values as \cite{Gorski2020}, thus ultimately concluding that this is not
a matter of some calibration error between OGLE-III and OGLE-IV data or
significant differences in methods of RC color determination.

We then compared our $E(V-I\,)$ reddening values (assuming an intrinsic
color gradient as determined in Sections~\ref{sec:vi0_lmc} 
and~\ref{sec:vi0_smc}) at the locations of eclipsing
binaries with $E(B-V\,)$ reddening values for these binaries estimated using
Na~I line \citep{Graczyk2018}, as was done by \cite{Gorski2020}. We convert
$E(B-V\,)$ to $E(V-I\,)$ with $E(V-I\,) = 1.318 \; E(B-V\,)$. The average
difference is $-0.040$ both in the LMC and SMC, such that our reddening
is lower than that of \cite{Graczyk2018}. This is fairly consistent
with the difference in the intrinsic color between \cite{Gorski2020} and this
work (0.045 and 0.048 for the LMC and SMC, respectively).
However, we found that the reddening of eclipsing binaries in the LMC estimated
by \cite{Graczyk2018} should not be averaged as it depends on the density of
stars, and so does the difference between reddening from \cite{Graczyk2018}
and this work.
Fig.~\ref{fig:compare-graczyk} shows the surface density of stars
in the centers of the LMC (panel~{\bf a}) and SMC (panel~{\bf c}). White
dots mark the locations of eclipsing binaries from \cite{Graczyk2018}.
Panels~{\bf b} and~{\bf d} show the change of the $E(V-I\,)$ difference
with the surface density of giant stars. In the case of the LMC (panel~{\bf b})
our reddening measurements are similar to those of \cite{Graczyk2018}
in low density areas, but become lower with growing surface density.
This can be explained in two ways: either there is a decrease of $(V-I\,)_0$
within the very center of the LMC bar, or there is a problem with a calibration
of the equivalent width of the Na~I line and the reddening.

In the latter case, the potential caveat is the use of an empirical relation
between gas and dust content in our Galaxy. This relation may be different
in the LMC and/or change with the gas metallicity, and thus could be the
source of error, although the detailed analysis of this problem is beyond our
area of expertise.

In the former case, we have already shown that there is an increase
of $(V-I\,)_0$ in the LMC, that results from the metallicity gradient.
The decrease of $(V-I\,)_0$ is possible for RC stars younger than 2~Gyr,
which have been excluded from our analysis.
However, we can compare reddening from RC stars with reddening from
another tracer to see if there is a similar behavior in the bar of the
LMC. If there is no dip in $E(V-I\,)$ in the LMC bar from another tracer,
then it would mean that our $E(V-I\,)$ values are underestimated, i.e. the
intrinsic RC color is overestimated. Fig.~\ref{fig:compare-rrlyr} shows such a
comparison between reddening of RC stars and RR Lyrae-type variable stars. The
RC sample (panel~{\bf a}) is presented in the lower resolution to match the one
of the RR Lyrae sample (panel~{\bf b}), where each bin contains an averaged
reddening value of at least 10 stars. The RR Lyrae data were taken from
\cite{Soszynski2016}, the photometric metallicity values from
\cite{Skowron2016}, and $(V-I\,)_0$ was calculated with the theoretical
metallicity-dependent relations from \cite{Catelan2004}. It is clearly visible,
that the central bar area with lower reddening is also present in the reddening
distribution of RR Lyrae stars. We therefore conclude, that this is caused by
the actual dust distribution in front of the bar region, and is not a result
of the change of intrinsic color.

In the case of the SMC, $(V-I\,)_0$ obtained by \cite{Gorski2020} is lower by
0.048~mag (or 0.031~mag if we exclude results from dubious tracers) than our
intrinsic color averaged in the area investigated by \cite{Gorski2020}.
Panels~{\bf c} and~{\bf d} in Fig.~\ref{fig:compare-graczyk}
show how the difference in reddening between \cite{Graczyk2014} and this work
depends on the density of RC stars. The sample of eclipsing binaries is very
small and does not span a sufficient range of stellar densities, so it is hard
to draw explicit conclusions. With the available data, it seems that $E(V-I\,)$
is almost independent of the surface density of stars.
This may be due to lower extinction in the SMC, lower star
densities and a small number of eclipsing binaries that could trace the
potential change of their reddening with star surface density.

\newpage

\section{Summary}
\label{sec:summary}

In this paper we present the most detailed and extensive $E(V-I\,)$ reddening
map of the Magellanic Clouds based on RC stars from the OGLE-IV survey
\citep{Udalski2015}, that provides reddening information for 180~deg$^2$ in the
LMC and 75~deg$^2$ in the SMC, with a resolution of $1.7' \times 1.7'$ in the
central parts of the MCs, decreasing to approximately $27' \times 27'$ in the
outskirts. The mean reddening in the LMC is $0.100 \pm 0.043$~mag in the LMC
and $0.047 \pm 0.025$~mag in the SMC.

We refine methods of calculating the RC color to obtain the highest possible
accuracy of reddening maps based on RC stars. In particular,
we account for the age-related additional RC structures that influence the
measurement of RC color.

From the spectroscopy of red giants \citep{Nidever2020} we find that there
is a small metallicity gradient with distance from the galaxy center in both
the LMC and SMC. Based on cluster data, we find a shallow dependence
of intrinsic RC color on cluster metallicity. By combining the two relations
we show that there is a slight decrease of the intrinsic RC color with
the distance from the center.
In the LMC $(V-I\,)_0 = 0.886 - d \times 0.002$, where $d$ is the distance
from the galaxy center. In the SMC $(V-I\,)_0 = 0.865 - d \times 0.003$
for $d>1.5^\circ$, and $(V-I\,)_0 = 0.877 - d \times 0.009$ for $d<1.5^\circ$,
where $d$ is the elliptical distance from the galaxy center.
This is consistent with the results of \cite{Nataf2020}, but contrary
to findings of \cite{Choi2018} and \cite{Gorski2020}.
We also argue that the low value of
$(V-I\,)_0$ obtained by \cite{Gorski2020} is a result of using unreliable
tracers, combined with possible calibration errors.

The reddening map is available on-line from the OGLE website at:

\vspace{0.1cm}
\hspace{0.5cm} {\it http://ogle.astrouw.edu.pl/cgi-ogle/get\_ms\_ext.py}
\vspace{0.1cm}

\noindent both for download (in TEXT and FITS formats, for the users convenience) and
in the form of an interactive interface.

The presented map covers the entire area where the RC is detectable and where
the spatial resolution it provides is better than that of the SFD reddening map,
thus exhausting the possibility of extending the detailed reddening map of the
MCs with the future RC data.

\acknowledgements
We thank Marek G\'{o}rski and David Nataf for fruitful discussions on
reddening in the MCs. We also thank the Referee for their valuable suggestions
that led to many improvements of this paper.
The OGLE project has received funding from the NCN grant MAESTRO
2014/14/A/ST9/00121 to AU.
DMS has received support from the NCN grant 2013/11/D/ST9/03445.\\
This research has made use of the NASA/IPAC Infrared Science Archive, which is
operated by the Jet Propulsion Laboratory, California Institute of Technology,
under contract with the National Aeronautics and Space Administration.\\
Funding for the Sloan Digital Sky Survey IV has been provided by the Alfred
P. Sloan Foundation, the U.S. Department of Energy Office of Science, and the
Participating Institutions. SDSS acknowledges support and resources from the
Center for High-Performance Computing at the University of Utah. The SDSS web
site is www.sdss.org.\\
SDSS is managed by the Astrophysical Research Consortium for the Participating
Institutions of the SDSS Collaboration including the Brazilian Participation
Group, the Carnegie Institution for Science, Carnegie Mellon University, Center
for Astrophysics | Harvard \& Smithsonian (CfA), the Chilean Participation
Group, the French Participation Group, Instituto de Astrof\'{i}sica de Canarias,
The Johns Hopkins University, Kavli Institute for the Physics and Mathematics
of the Universe (IPMU) / University of Tokyo, the Korean Participation Group,
Lawrence Berkeley National Laboratory, Leibniz Institut f\"{u}r Astrophysik
Potsdam (AIP), Max-Planck-Institut f\"{u}r Astronomie (MPIA Heidelberg),
Max-Planck-Institut f\"{u}r Astrophysik (MPA Garching), Max-Planck-Institut
f\"{u}r Extraterrestrische Physik (MPE), National Astronomical Observatories
of China, New Mexico State University, New York University, University of
Notre Dame, Observat\'{o}rio Nacional / MCTI, The Ohio State University,
Pennsylvania State University, Shanghai Astronomical Observatory, United
Kingdom Participation Group, Universidad Nacional Aut\'{o}noma de M\'{e}xico,
University of Arizona, University of Colorado Boulder, University of Oxford,
University of Portsmouth, University of Utah, University of Virginia,
University of Washington, University of Wisconsin, Vanderbilt University,
and Yale University.\\
This work has made use of NASA's Astrophysics Data System Bibliographic
Services.


\end{document}